\documentclass[11pt]{article}

\usepackage{amsthm,amsmath,amsfonts,amssymb,booktabs,graphicx,url}
\usepackage[authoryear]{natbib}
\usepackage[margin = 3.3cm]{geometry}

\newtheorem{prop}{Proposition}

\begin{document}

\title{Prediction intervals for economic fixed-event forecasts\thanks{We thank two anonymous reviewers, Andreas Eberl, Tilmann Gneiting, Malte Kn\"uppel as well as participants of the MathSEE symposium (Karlsruhe, September 2023) and seminar participants at the Universities of Amsterdam and Freiburg for helpful comments. We further thank Alexander Henzi for sharing program code associated with \cite{Henzi2022}, and Eben Lazarus for sharing code related to \cite{LazarusEtAl2018}. We acknowledge support by the state of Baden-W\"urttemberg through bwHPC.}}

\author{Fabian Kr\"uger \\ Karlsruhe Institute of Technology \and Hendrik Plett \\ ETH Z\"urich}

\maketitle

\begin{abstract}
The fixed-event forecasting setup is common in economic policy. It involves a sequence of forecasts of the same (`fixed') predictand, so that the difficulty of the forecasting problem decreases over time. Fixed-event point forecasts are typically published without a quantitative measure of uncertainty. To construct such a measure, we consider forecast postprocessing techniques tailored to the fixed-event case. We develop regression methods that impose constraints motivated by the problem at hand, and use these methods to construct prediction intervals for gross domestic product (GDP) growth in Germany and the US. 
\end{abstract}

\section{Introduction}

Economic forecasts are often published in the `fixed-event' format. Consider \cite{tagesschau2022}, a German news website that maintains a list of recent GDP forecasts made by various institutions. For example, in April 2022, the federal government predicted 2.2\% GDP growth for 2022, followed by 2.5\% in 2023. In July 2022, the European Commission predicted 1.4\% growth for 2022, followed by 1.3\% in 2023. These forecasts are called `fixed-event' because the quantity of interest -- the GDP growth rate over the year 2022 or 2023 -- remains fixed, whereas the date at which the forecast is made moves forward in time. This forecasting format is routinely used by institutions that make economic forecasts, and by news media which comment on these forecasts.

Economic fixed-event forecasts are typically published without a quantitative measure of uncertainty. Such a measure would be particularly important in the current setup: By construction, forecast uncertainty varies heavily depending on when the forecast is made and which year it refers to. For example, in July 2022, the EU commission's forecast for 2022 should be less uncertain than its forecast for 2023. But just how much less uncertain depends on the time series properties of GDP, and is hard to grasp intuitively.

Motivated by this situation, the present paper constructs measures for the uncertainty in fixed-event point forecasts. Together with the point forecasts themselves, we can then construct forecast distributions for GDP growth and other economic variables. The principle of forecast post-processing -- using point forecasts and past forecast errors to construct forecast distributions -- is popular in meteorology (see e.g. \citealt{GneitingRaftery2005}, \citealt{GneitingEtAl2005}, \citealt{RaspLerch2018} and \citealt{Vannitsem2021}), economics (e.g. \citealt{Knueppel2014}, \citealt{KruegerNolte2016} and \citealt{Clark2020}) and other fields. State-of-the-art point forecasts are often publicly available, so that using them as a basis for forecast distributions is more practical than generating forecast distributions from scratch. Furthermore, assessing forecast uncertainty based on past errors does not require knowledge about how the point forecasts were generated. This is an important advantage in practice, where the forecasting process may be judgmental, subject to institutional idiosyncracies, or simply unknown to the public.

The vast majority of the postprocessing literature considers a `fixed-horizon' forecasting setup where the time between the forecast and the realization remains constant. Examples of the fixed-horizon case include daily forecasts of temperature 12 hours ahead, or quarterly forecasts of the inflation rate between the current and next quarter. In economics, \cite{Clements2018} constructs a measure of fixed-event forecast uncertainty that is based on fixed-horizon forecast errors, and thus requires that an appropriate database of fixed-horizon forecasts is available. This is the case in the US Survey of Professional Forecasters analyzed by \citeauthor{Clements2018}, but not in other situations including the German GDP example mentioned earlier. Existing approaches are thus not applicable to the fixed-event case. Instead, the latter requires different tools which we develop in this paper.

The main idea behind our proposed approach is simple: We model quantiles of the forecast error distribution as a function of the forecast horizon. The latter is defined as the time (measured in weeks) between the forecast and the end of the target year. For example, forecasts made on July 1, 2022 for 2022 and 2023 correspond to horizons $26$ and $78$, respectively. To estimate regression models, we use a dataset of past forecast errors at different horizons. Pooling forecast errors across horizons allows us to estimate uncertainty at any given horizon by considering uncertainty at neighboring horizons. This approach is helpful in the fixed-event case, where only a small number of past errors is typically available for a given horizon. For example, the German data set we consider covers forecast-observation pairs ranging from 1991 to 2022, and includes precise information (daily time stamps) on the forecast horizon $h$. Specifically, the data contains $525$ unique values for $h$, ranging from $h = 0$ to $h = 104$. Note that $h$ need not be an integer; for example, forecasts made on the seven days of the target year's final week correspond to horizons $h \in \{0, 1/7, \ldots, 6/7\}$. Given that the data covers $n = 1\,307$ observations in total, the average number of forecast errors corresponding to each of the $525$ different horizons is about $2.5$. Using only forecast errors that correspond exactly to some horizon of interest ($h = 5$ weeks, say) is hence not a promising strategy, and considering forecast errors from neighboring horizons seems advisable. This aspect is not relevant when postprocessing fixed-horizon forecasts, which is typically based on a time series of past forecast errors for the exact horizon of interest.

The statistical methods we consider incorporate constraints that are motivated by the fixed-event forecasting problem. In particular, it is plausible to assume that forecast uncertainty increases monotonically across horizons, and levels off at some horizon. We consider three main approaches that implement this idea: A Gaussian heteroscedastic model, a decomposition approach that imposes a symmetry assumption, and a flexible approach that is (almost) nonparametric. The first two approaches are parsimonious in light of typically short samples of past forecast errors. Despite their simplicity, they perform well in a cross-validated analysis of German and US data. In particular, the prediction intervals attain coverage close to its nominal level, and they clearly outperform benchmark predictions by a survey of professional economists for the US data.

To illustrate the quantitative implications of our results, consider a hypothetical forecaster who, as of mid-September, issues a point prediction of 2.1\% GDP growth for the current year, and 1.7\% for the next year. Our results for the German data then suggest that a plausible 80\% prediction interval for the current year would be on the order of $2.1 \pm 0.35 = [1.75,2.45]$, compared to $1.7 \pm 2 = [-.3, 3.7]$ for the next year, i.e., the prediction interval for the next year is more than five times as wide. We thus argue that common statements along the lines of `we expect 2.1\% GDP growth this year, and 1.7\% next year' are not helpful: They implicitly put the current- and next-year forecasts on an equal footing, and over-emphasize the next-year point forecast which is surrounded by considerable uncertainty.

The rest of this paper is structured as follows. Section \ref{sec:setup} introduces notation for the fixed-event setup. In order to illustrate the statistical implications of this setup, Section \ref{sec:ar1model} considers fixed-event forecasting in an autoregressive time series model. Section \ref{sec:models} introduces the proposed regression methods for forecast postprocessing. Section \ref{sec:eval} describes methodology for forecast evaluation and model selection. Section \ref{sec:sim} studies the performance of the proposed methods in a simulation experiment, and Section \ref{sec:emp} considers empirical fixed-event forecasts from Germany and the US. Section \ref{sec:disc} concludes with a discussion. The online supplement contains derivations and further analyses. Replication materials are available at \url{https://github.com/FK83/gdp_intervals}.

\section{Setup}\label{sec:setup}

We consider forecasting $Y_t$, the real GDP growth rate in year $t$. The latter is defined as $100 \times \frac{\text{GDP}_t-\text{GDP}_{t-1}}{\text{GDP}_{t-1}},$ where $\text{GDP}_t$ is the level of real GDP in year $t$, which in turn is taken to be the average of the year's four quarterly levels. For many countries, point forecasts of $Y_t$ are available from various public and private forecasting institutions. We denote the time between the origin date (on which the forecast is issued) and the end of year $t$ as the `forecast horizon', denoted by the symbol $h$ and measured in weeks. Let $X_{t,h}$ denote the point forecast of $Y_t$ at horizon $h$. For example, for year $t = 2020$, a forecast issued on December 17, 2020 corresponds to horizon $h = 2$. We focus on forecast horizons $h \in [0, 104]$, i.e., up to two years ahead, which covers most practical economic forecasts. Even a forecast at horizon $h = 0$ need not be perfect in practice since a precise estimate of the outcome may be available only after the end of year $t$. For simplicity, we treat each year as having $52$ weeks.

The forecast error of $X_{t,h}$ is given by $e_{t,h} = Y_t-X_{t,h}$. Let $\mathcal{F}_{t,h}$ denote a suitable information set that is available $h$ weeks before the end of year $t$. 
In order to obtain a forecast distribution for $Y_t$ given $\mathcal{F}_{t,h}$, or $Y_t|\mathcal{F}_{t,h}$ in short, we 
model the distribution of $e_{t,h}|\mathcal{F}_{t,h}$. That is, we take the point forecast $X_{t,h}$ as given, and focus on modeling the error of this point forecast. Since $$\mathbb{P}(Y_t \le y|\mathcal{F}_{t,h}) = \mathbb{P}(e_{t,h} \le (y-X_{t,h})|\mathcal{F}_{t,h}),$$ and $X_{t,h}$ is known given $\mathcal{F}_{t,h}$, the distribution $e_{t,h}|\mathcal{F}_{t,h}$ can be used to construct the desired distribution for $Y_t|\mathcal{F}_{t,h}$.

Our approach of modeling the error of a given point forecast, rather than constructing a forecast distribution from scratch, is called `postprocessing'.\footnote{In meteorology, the term `ensemble postprocessing' is more common since postprocessing is typically applied to a collection (`ensemble') of point forecasts stemming from a numerical weather prediction model \citep[see][]{GneitingRaftery2005}. That said, the broader principle of modeling forecast errors readily transfers to the case of a single point forecast that we consider here.} However, all postprocessing studies we are aware of consider fixed-horizon forecasting. Our fixed-event setup requires different statistical tools in that only a small number of observations is typically available for a given forecast horizon. We thus focus on modeling the properties of $e_{t,h}$ as a continuous function of $h$, thus interpolating across forecast horizons. By contrast, postprocessing methods for fixed-horizon forecasts typically treat each horizon separately, using a time series $(e_{t,h})_{t=1}^n$ of past forecast errors at a single horizon $h$. 

\section{Illustrating fixed-event forecasting via an autoregressive time series model}\label{sec:ar1model}

In this section, we consider fixed-event forecasting in an autoregressive Gaussian time series model from the econometric literature. The stylized facts illustrated here will later motivate our more general empirical methodology (described in Section \ref{sec:models}). 

\subsection{Autoregressive model for weekly GDP growth}

Our model is a modified version of the one in \cite{PattonTimmermann2011}. It assumes that fixed-event forecasts are based upon noisy high-frequency observations, but make correct use of these observations. That is, forecasts are equal to the true conditional mean of the predictand, given their information base. While high-frequency data on GDP is not literally available in practice (where GDP is measured at a quarterly frequency only), there are various efforts at measuring economic activity based on economic variables that are available at monthly, weekly or daily frequency \citep{AruobaEtAl2009,BraveEtAl2019,LewisEtAl2021,EraslanGoetz2021}. Furthermore, publication lags and ex-post revisions in macroeconomic data \citep[see e.g.][]{Croushore2001} imply that realizations become available with a delay. Taken together, the model's assumption of a noisy proxy observed at high frequency hence provides a plausible (albeit abstract) representation of practical GDP forecasting. 

We consider hypothetical weekly observations, whereas \citeauthor{PattonTimmermann2011} use hypothetical monthly observations. This increase in granularity is motivated by our empirical data setup, where many forecasts are made within the month, so that a monthly frequency may be too coarse. Section A.1 of the online supplement provides a detailed comparison of our model to the one of \citeauthor{PattonTimmermann2011}. We approximate $Y_t$, the GDP growth rate from year $t-1$ to year $t$, as the weighted sum of $103$ weekly logarithmic growth rates ranging from the beginning of year $t-1$ to the end of year $t$. As detailed in the online supplement, this setup arises from the definitorial convention ('annual-average') that we use to compute $Y_t$. We denote the weekly logarithmic growth rate by $Y_w^*$, with the understanding that each year $t$ corresponds to a distinct set of 52 index values $w$ (see below for an example). Here and henceforth, we use the `star' superscript notation for weekly random variables. We further assume that $Y_w^*$ follows a first-order autoregression, so that
\begin{eqnarray}
	Y_w^* &=& \rho~Y^*_{w-1} + \varepsilon_w^* \label{arw1}\\
	\varepsilon_w^* &\stackrel{\text{iid}}{\sim}& \mathcal{N}(0, \sigma^2_\varepsilon)\\
	Y_t &\approx& \sum_{j=1}^{103} \gamma_{j}~Y_{w_\text{last}(t)+1-j}^*, \label{arw3}
\end{eqnarray}
where $w_\text{last}(t)$ denotes the index of the last week of year $t$, and $\gamma_j = 1-\frac{|52-j|}{52}$. Note that the coefficients $\gamma_j$ form a triangle when plotted against $j$. For example, suppose that the sample starts in year $t = 1$ (containing weeks $w = 1, 2, \ldots, 52$), so that week $w = 104 = w_\text{last}(2)$ is the last week of year $2$. According to Equation (\ref{arw3}), the approximate annual growth rate $Y_2$ is a weighted sum of the weekly observations $(Y_w^*)_{w=2}^{104}$. The greatest weight of $1$ is associated with $Y^*_{53}$, and the smallest weight of $1/52$ is associated with $Y^*_{104}$ and $Y^*_{2}$. Furthermore, $\sum_{j=1}^{103} \gamma_j = 52,$ in line with the fact that $Y_w^*$ is measured at a weekly frequency whereas $Y_t$ is measured annually.
See Section A.2 of the online supplement for a concise derivation of the approximation at (\ref{arw3}), and \citet[Appendix B]{PattonTimmermann2011} for numerical evidence on the high precision of the approximation. 

As noted earlier, we assume that forecasters observe a noisy version of $Y_w^*$. Specifically, let $\tilde Y_w^* = Y_w^* + \eta_w^*,$ where $\eta_w^* \stackrel{\text{i.i.d.}}{\sim} \mathcal{N}(0,\sigma^2_\eta)$ is an independent and identically distributed (IID) Gaussian measurement error. An $h$-week ahead mean forecast of $Y_t$ is then based on the information set $\mathcal{F}_{t,h}$ generated by the sequence of noisy weekly observations $(\tilde Y_w^*)_{w=1}^{w_\text{last}(t)-h}.$ Due to the presence of measurement error, the optimal forecast of $Y_t$ given $\mathcal{F}_{t,h}$ has no simple closed-form expression. However, the optimal forecast can be computed analytically by means of the Kalman filter. The latter uses the model's linear state space representation, which we describe in Section A.3 of the online supplement.

An important implication of the present model is that forecasts of $Y_t$ at horizon $h = 0$ are not perfect, which is in line with reality. For example, the GDP for 2023 is not yet published on December 31, 2023, so that the $0$-week ahead forecast of $Y_t$ is formed on the basis of preliminary estimates.

\begin{figure}
	\includegraphics[width=\textwidth]{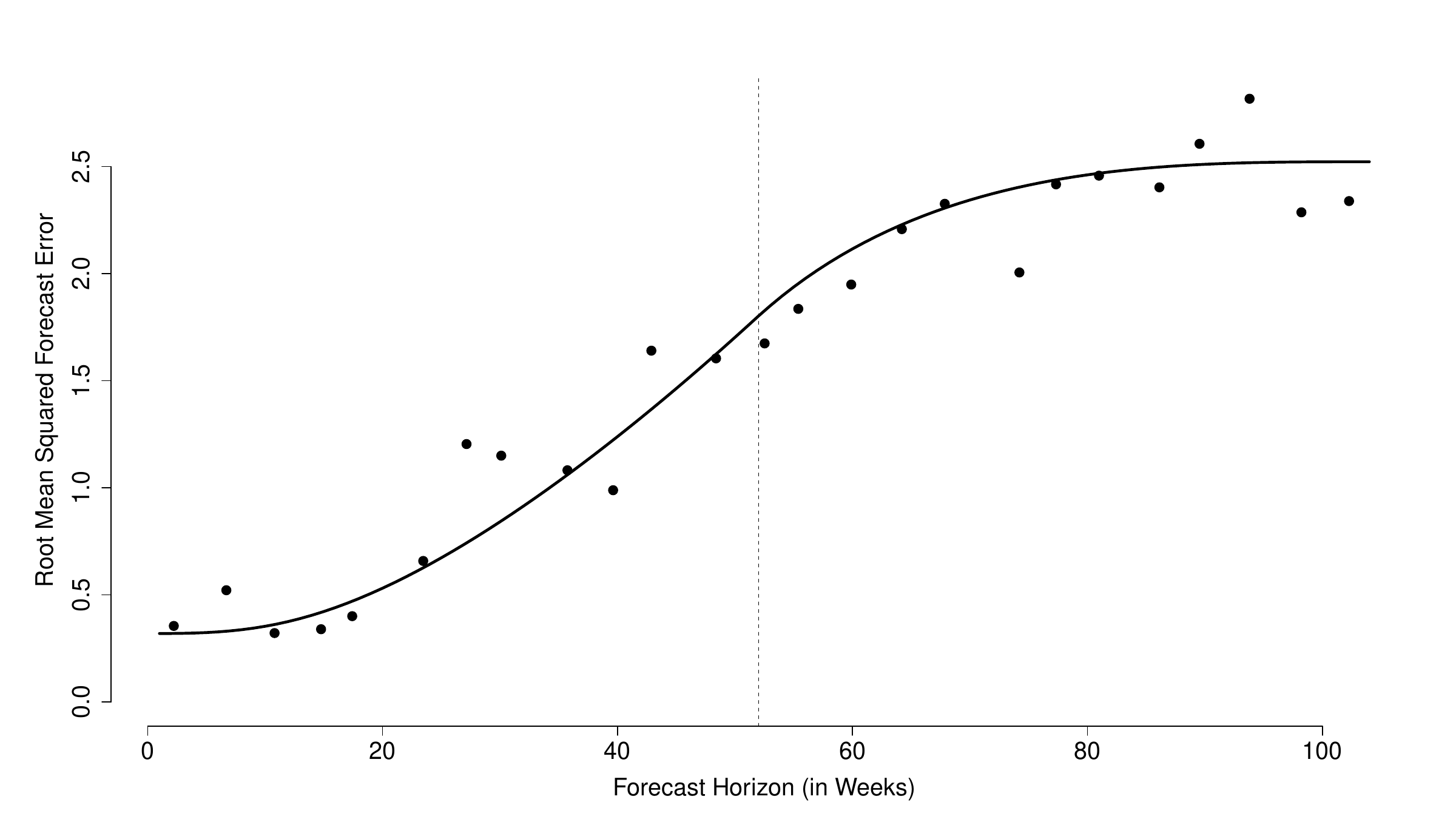}
	\caption{The curve shows model-implied root mean squared forecast errors for $\rho = 0.3$, $\sigma^2_\varepsilon = 0.09,$ and $\sigma^2_\eta = 0.003$. Dots show empirical root mean squared errors (estimated based on $26$ disjoint bins for $h$) for the German GDP data studied in Section \ref{sec:emp}. The forecast horizon of one year ($52$ weeks) is marked by a vertical line.\label{fig:rmse}}
\end{figure}

\subsection{Evolution of forecast uncertainty}\label{sec:term}

The model allows to study the root mean squared forecast error (RMSFE) of fixed-event point forecasts as a function of the forecast horizon. This relationship represents the amount of predictability (or lack thereof) at various points in time, and will be relevant for specifying appropriate postprocessing approaches below. Figure \ref{fig:rmse} presents the relationship. We have set the three model parameters $\rho = 0.3$ (persistence of weekly GDP growth), $\sigma^2_\varepsilon = 0.09$ (variance of white noise error in weekly GDP growth) and $\sigma^2_\eta = 0.003$ (variance of measurement error) that yield a plausible approximation of the empirical RMSFE values observed in the German GDP data that we consider below. The figure shows that the model-implied RMSFE increases as the forecast horizon increases. This type of monotonicity is a well-known property of forecasts that are optimal given some information set \citep[see e.g.][]{Patton2012,Krueger2021}, which is satisfied here. The shape of the function further indicates a rather steep and roughly linear increase in RMSFEs between $h \approx 20$ and $h \approx 70$, whereas RMSFEs are almost constant for $h \le 10$ and $h \ge 80$. In Section A.4 of the online supplement, we explore the RMSFE curves implied by other values of the model's parameters.

\subsection{Samples of past forecast errors}\label{sec:sample}

Our empirical analysis below is based on a sample of past forecast errors $(e_{t_i, h_i})_{i=1}^n$, where $t_i$ and $h_i$ denote the target year and horizon of the $i$th forecast error. We next discuss the correlation structure of such a sample in the context of the model considered above. For simplicity, we assume that $\sigma^2_\eta = 0$, such that weekly data $Y_w^*$ are observed without error.

Consider two forecast error observations $e_{t_1, h_1}$ and $e_{t_2, h_2}$. Without loss of generality, let $t_2 \ge t_1$. Furthermore, we focus on the case $1 \le h_2 \le 104$ (forecast errors ranging from one week to two years). The two forecast errors are independent if one of the following sufficient conditions holds:
\begin{itemize}
	\item $t_2 = t_1 + 1, h_2 \le 52,$ or
	\item $t_2 \ge t_1 + 2$; 
\end{itemize}
see Section A.5 of the online supplement for details. Intuitively, both conditions rule out any overlap between the two time intervals  $[t_1-h_1, t_1]$ and $[t_2-h_2, t_2]$ that range from the forecast date to the target date of the two forecast errors. If neither of the conditions holds, the two forecast errors will typically be dependent.

More broadly, the dependence structure in a sample of forecast errors $(e_{t_i, h_i})_{i=1}^n$ thus reveals a cluster-type pattern. If $h \le 52$, then the forecast errors belonging to a given year (e.g., all errors $e_{t_i, h_i}$ s.t. $t_i = 2019$) form one cluster, with errors being dependent within the same cluster but independent across different clusters. If $h_i \ge 53$ is possible, as is the case in our empirical analysis, then the dependence structure is more complicated, with possibly nonzero correlation across neighboring years. If we allow for the case $\sigma^2_{\eta} > 0$, the dependence structure becomes even more complicated since any pair of forecast errors may be jointly affected by updated `back-casts' (i.e., corrected assessments of past data, as produced by the Kalman filter) that are relevant in the presence of measurement error. 

\cite{HansenLee2019} present asymptotic results on samples with cluster dependence, and on estimators based upon such samples. Similar results can be derived in the present setup, and allow to derive asymptotic statements as the number of clusters tends to infinity. However, in view of the rather short data samples used in our empirical analysis (about $32$ years for the German data, and about $42$ years for the US data), it is unclear whether such results provide a good description of our setup. In Section \ref{sec:sim}, we hence conduct a simulation study that closely mimics our empirical setup.

\section{Modeling the distribution of forecast errors}\label{sec:models}

The model discussed in Section \ref{sec:ar1model} is useful to illustrate the main stylized facts of the fixed-event forecasting problem. \cite{PattonTimmermann2011} estimate the model's parameters and use the model for predicting the distribution of forecast errors as a function of $h$. While conceptually appealing, estimation is challenging in practice, with parameter estimates differing markedly across estimation methods \citep[see][Table 2]{PattonTimmermann2011}. Addressing these challenges seems unnecessary in the current setup, where interest lies on forecasting (as opposed to interpreting the model's structural parameters). In the following, we therefore focus on empirical models that are considerably simpler to implement and are less restrictive in terms of functional form assumptions. In the simulation experiments from Section \ref{sec:sim}, we demonstrate that these empirical methods perform well even when the true data-generating process is given by the model from Section \ref{sec:ar1model}.

To simplify model building, we assume that the distribution of $e_{t,h}|\mathcal{F}_{t,h}$ is a function of $h$ alone, i.e. that $\mathbb{P}(e_{t,h} \le z|\mathcal{F}_{t-h}) = \int_{-\infty}^z dF_h(z),$ where $F_h$ is a distribution that depends on $h$ but is constant over time. This assumption is satisfied, for example, for the steady state version of the autoregressive model from Section \ref{sec:ar1model}.\footnote{Here `steady state' means that a sequence of initial observations has been removed, so that the prior mean and variance for the initial state vector become irrelevant.\label{fn:steady}} From a pragmatic perspective, the assumption of a time-invariant conditional distribution function is motivated by the paucity of data in our empirical analysis, which implies that estimating elaborate conditional distributions does not seem promising. However, we retain the dependence on $h$, which we expect to be a major determinant of forecast error distributions, with larger values of $h$ corresponding to more variable forecast errors.

In Sections \ref{sec:gauss} to \ref{sec:flexible}, we describe three modeling approaches, all of which incorporate constraints that are motivated by the fixed-event forecasting setup. On the other hand, the stringency of the imposed constraints differs across approaches.

\subsection{Gaussian approach}\label{sec:gauss}

\begin{figure}[!h]
	\includegraphics[width=\textwidth]{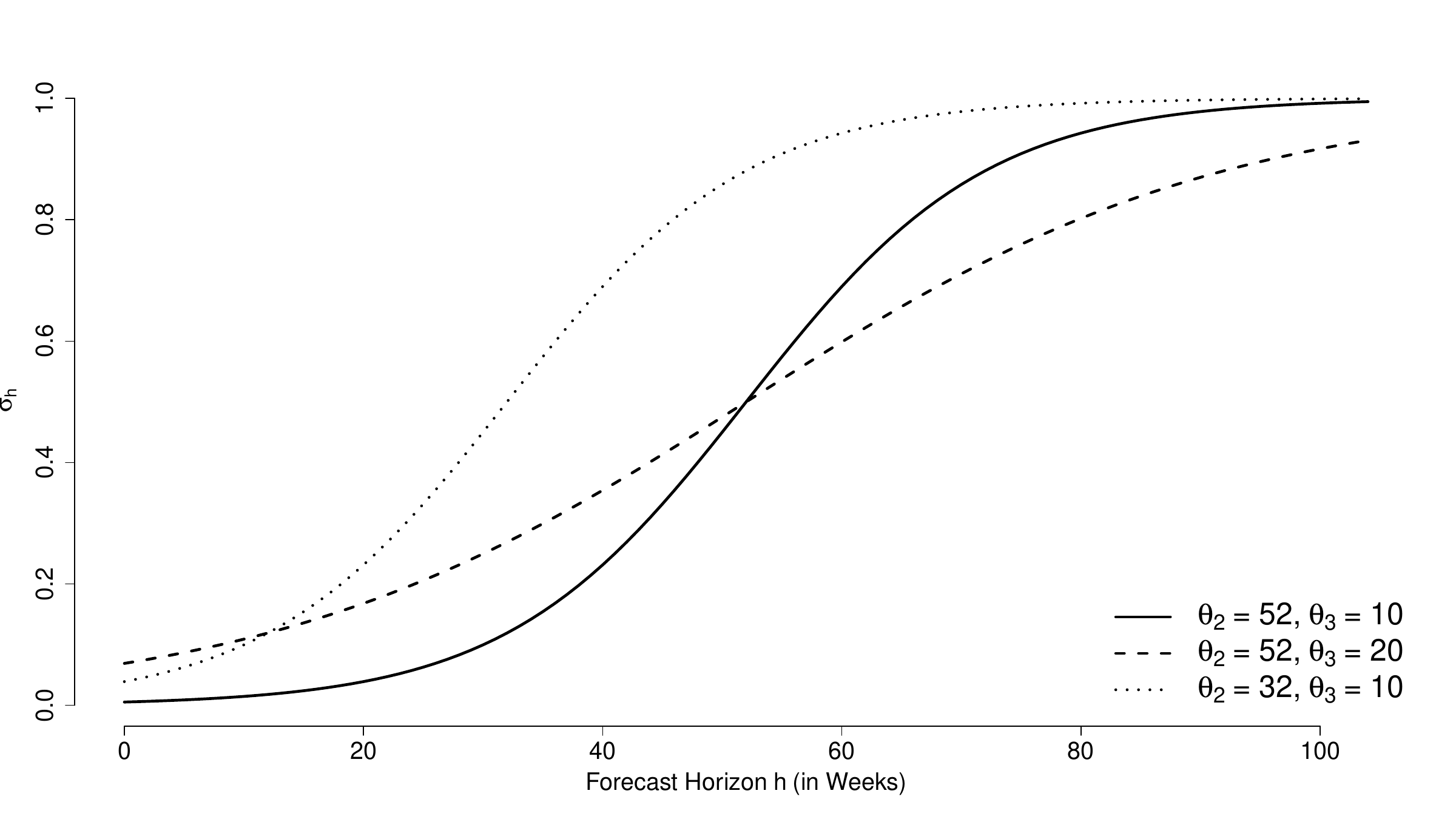}
	\caption{Illustration of the logistic function for $\sigma_h$ in Equation (\ref{eqn:sigmah}). As described in the legend (bottom right), the three curves refer to different paramter choices for $\theta_2$ and $\theta_3$. The parameter $\theta_1$ is set to one in each case.\label{fig:illustrate_logistic}}
\end{figure}

Our first, most restrictive approach assumes that $e_{t,h} \sim \mathcal{N}(\mu, \sigma_h^2),$ with
\begin{equation}
	\sigma_h = \theta_1 \times \frac{1}{1 + \exp\left(-\frac{(h-\theta_2)}{\theta_3}\right)};\label{eqn:sigmah}
\end{equation}
note that the second factor in (\ref{eqn:sigmah}) is the cumulative distribution function (CDF) of a logistic random variable with location $\theta_2$ and scale $\theta_3 > 0$, evaluated at $h$. We hence model the standard deviation of $e_{t,h}$ as a constant ($\theta_1 > 0$) times a function that is monotonically increasing in $h$. As noted above, the monotonicity of $\sigma_h$ is an implication of optimal forecasting. In practice, we expect monotonicity to hold also under moderate forms of sub-optimality. The logistic functional form of $\sigma_h$ is motivated by the structural model displayed in Figure \ref{fig:rmse}. The specification at (\ref{eqn:sigmah}) implies that
$\sigma_h \rightarrow \theta_1 \equiv \sigma_\infty$ as $h \rightarrow \infty$, and $\sigma_h \rightarrow 0$ as $h \rightarrow -\infty$. Both implications seem economically plausible, noting that negative horizons ($h < 0$) correspond to back-casts that, for sufficiently small values $h << 0$, are no longer affected by data revisions or missing input data. The parameter $\theta_2$ denotes the horizon $h^{'}$ that satisfies $\sigma_{h^{'}} = 0.5~\sigma_\infty$, i.e., the horizon at which half of the uncertainty is resolved, whereas $\theta_3$ is a shape parameter that provides further flexibility. Figure \ref{fig:illustrate_logistic} illustrates the functional form of $\sigma_h$ for different parameter choices. The mean parameter $\mu$ is a parsimonious way to allow for nonzero forecast errors resulting from biased point forecasts. While more sophisticated specifications are possible (with the bias depending, for example, on the forecast horizon $h$), the empirical results in  Section \ref{sec:emp} (in particular, Table S2 in the online supplement) indicate that the empirical importance of the bias is quite limited in our setup. 

We estimate the Gaussian model by minimizing the continuous ranked probability score \citep[CRPS;][]{MathesonWinkler1976}, as implemented in the scoringRules software package \citep{JordanEtAl2019} for R \citep{R}. The CRPS is a loss function for distributions, and has become a popular alternative to the log likelihood function (also called logarithmic score) in recent years. The CRPS is a strictly proper scoring rule, that is, a forecaster has an incentive to state what they think is the true forecast distribution. A general advantage of the CRPS over the logarithmic score is that it does not require a density, and can easily handle discrete or empirical distributions \citep[see][]{JordanEtAl2019,KruegerEtAl2021}. When used as a criterion for estimation, the CRPS yields consistent parameter estimates under standard conditions including correct specification  \citep{GneitingRaftery2007}. We use the CRPS for parameter estimation since it is related to the interval score that we use for forecast evaluation (see Section \ref{sec:eval} for details), thus broadly aligning the criteria used for model estimation versus evaluation.

\subsection{Decomposition approach}

Our second model is based on a decomposition of $e_{t,h}$ into its sign and its (absolute) magnitude. Similar decompositions have been considered for modeling financial returns \citep{ChristoffersenDiebold2006,AnatolyevGospodinov2010}, based on the motivation that the sign of returns is far less predictable than their magnitude. Noting that a similar motivation applies to forecast errors, we adopt a similar decomposition method here. To describe the method, note that
$$e_{t,h} = \underbrace{(2~\mathbf{1}(e_{t,h} > 0)-1)}_{=S_{t,h}}~|e_{t,h}|,$$
where $\mathbf{1}(A)$ is the indicator function of the event $A$. We assume that the sign $S_{t,h}$ takes values of $\pm 1$ with equal probability, and is independent of the magnitude $|e_{t,h}|$. We further assume that the CDF of $|e_{t,h}|$, which we denote by $G_h$, is constant across time $t$ but (possibly) different for each forecast horizon $h$. Under these assumptions, we obtain
\begin{eqnarray}
	\mathbb{P}(e_{t,h} \le z|\mathcal{F}_{t-h}) &=& \begin{cases} 0.5 + 0.5~G_h(|z|) &~\text{if}~ z > 0 \\  0.5 - 0.5~G_h(|z|) &~\text{if}~ z \le 0.  \end{cases} \label{decomp}
\end{eqnarray}
Finally, we assume that $|e_{t,h}|~{\succsim}_{\text{FSD}}~ |e_{t,h-v}|$ for any $v > 0$, where the notation ${\succsim}_{\text{FSD}}$ indicates first-order stochastic dominance (FSD). This means that the distribution of absolute forecast errors becomes stochastically greater as the horizon increases. This assumption seems plausible in the present context, and can be motivated as a more stringent version of increasing mean squared forecast errors across horizons.\footnote{Specifically, the assumption implies that $e_{t,h}~{\succsim_\text{CX}}~e_{t,h-v}$ for $v > 0$, where $\succsim_\text{CX}$ denotes convex order \citep[see][Section 3.D]{ShakedShanthiku2007}. The latter relation implies that $\mathbb{E}(e_{t,h}^2) \ge \mathbb{E}(e_{t,h-v}^2),$ i.e., increasing mean squared forecast errors.} In order to estimate $G_h$ under the present assumption, we use isotonic distributional regression (IDR) as recently studied by \cite{HenziEtAl2021} and implemented in the R package isodistrreg \citep{HenziEtAl2022}. Briefly, IDR estimates a conditional distribution function, subject to the constraint that the outcome increases (in the sense of FSD) as the predictor vector increases (with various notions of `increases' being covered, including partial orders on the covariate space). Our empirical setup is a fairly simple special case, in that we use a single continuous predictor ($h$). See \citet[Equation 1 and Figure 1]{HenziEtAl2021} for an instructive example of IDR in this case. IDR has two main advantages over unconstrained nonparametric estimators of the distribution of $|e_{t,h}|$ given $h$. First, imposing the assumption serves to regularize the estimator and reduce estimation noise. Second, IDR is free of tuning parameters, whereas bandwidth parameters are often crucial for the performance of conventional nonparametric estimators. Note that for a given horizon $h$, Equation (\ref{decomp}) implies that a forecast distribution for $e_{t,h}$ obtained via the decomposition approach is necessarily symmetric around zero.

\subsection{Flexible approach}\label{sec:flexible}

Our most flexible model estimates $F_h(e_{t,h})$ nonparametrically, subject to the constraint that $e_{t,h}~{\succsim}_{\text{ICX}}~ e_{t,h-v}$ for any $v > 0$, where ICX denotes increasing convex order. For two random variables $V$ and $W$, $V~ {\succsim}_{\text{ICX}}~W$ holds if $\mathbb{E}(\phi(V)) \ge \mathbb{E}(\phi(W))$ for all increasing convex functions $\phi$. Imposing the ICX constraint is hence less restrictive than imposing either FSD (which requires that the inequality holds for all increasing functions $\phi$) or convex order (which requires that the inequality holds for all convex functions $\phi$). To estimate the distribution of $e_{t,h}$ under the ICX constraint, we use the recent approach of \cite{Henzi2022}. Similar to IDR, a major advantage of this approach is that it does not rely on tuning parameters. Section 1 of \cite{Henzi2022} provides further discussion of the ICX constraint, whereas Sections 5 and 6 present simulation and empirical examples. In particular, Equation (2) of \cite{Henzi2022} illustrates a data-generating process that satisfies ICX but violates FSD. Note that for a given horizon $h$, a forecast distribution for $e_{t,h}$ obtained via the flexible approach need not be symmetric around zero (or around any other value). This is because the approach does not assume a particular shape of the distribution $F_h$ for any given $h$. Instead, it assumes that two distributions $F_{h_1}$ and $F_{h_2}$ are ordered according to ICX. 

\subsection{Comparison}

The following result establishes that, in terms of the respective assumptions, the Gaussian modeling approach with $\mu = 0$ is a special case of the decomposition approach which, in turn, is a special case of the flexible approach. 

\begin{prop}
	\begin{itemize}
		\item[(a)] Let $e_{t,h} \sim \mathcal{N}(0, \sigma^2_h)$, where $\sigma^2_h$ is increasing in $h$. Then the sign of $e_{t,h}$ takes values $\pm 1$ with equal probability and is independent of $|e_{t,h}|$. Furthermore, $|e_{t,h}|$ is stochastically increasing in $h$.
		\item[(b)] If $e_{t,h}$ is generated according to the decomposition model described around Equation (\ref{decomp}), then it satisfies increasing convex order as $h$ increases. 
	\end{itemize}
\end{prop}
\begin{proof}
	\begin{itemize}
		\item[(a)] For the first part, note that a standard normal variate $Z$ satisfies $\mathbb{P}(|Z| \ge b|Z > 0) = \mathbb{P}(|Z| \ge b|Z \le 0) = 2~(1-\Phi(b)),$ where $b > 0$. Hence for any set $B \subset \mathbb{R}_+$, it holds that $\mathbb{P}(|Z| \in B|Z > 0) = \mathbb{P}(|Z| \in B|Z \le 0)$. Independence of $|Z|$ and $\mathbf{1}(Z > 0)$ then follows immediately. For the second part, note that $\mathbb{P}(|e_{t,h}| \le b) = \mathbb{P}(-b \le e_{t,h} \le b) = 2~\Phi(b/\sigma_h)-1,$ which is decreasing in $h$. Hence $|e_{t,h}|$ is stochastically increasing in $h$. 
		\item[(b)] Under the assumptions of the decomposition model, $|e_{t,h}|$ is stochastically increasing in $h$ (in the sense of FSD). The latter implies convex ordering of $e_{t,h}$ \citep[see][Section 3.D]{ShakedShanthiku2007}, which in turn implies increasing convex ordering of $e_{t,h}$.
	\end{itemize}\vspace{-.4cm}
\end{proof}
In its steady-state version (see Footnote \ref{fn:steady}), the autoregressive model from Section \ref{sec:ar1model} satisfies the assumptions of the proposition's part (a). More specifically, the forecast at horizon $h$ is optimal given a certain information set (consisting of all noisy observations up to the forecast date).  Together with the model's Gaussian setup, this implies that forecast errors $e_{t,h}$ are Gaussian with mean zero and variance $\sigma^2_h$, where the latter is increasing in $h$ \citep[see, e.g.,][Section 2.2]{Patton2012}. 

\subsection{Combination}\label{sec:comb}

In addition to the three postprocessing methods introduced above, we consider a simple combination method: For a given quantile level (10\% or 90\%), we use the arithmetic mean of the three methods' forecasts. From a practical perspective, a main appeal of combinations is that their good performance is somewhat predictable, based on a large body of empirical evidence and theoretical findings on the properties of combinations \citep[e.g.][]{GneitingRanjan2013,Lichtendahl2013,WangEtAl2022}. By contrast, predicting the relative performance of several forecasting methods is typically more difficult in practice.

\section{Forecast evaluation}
\label{sec:eval}

\subsection{Measure of forecast accuracy}

We focus on prediction intervals, and evaluate forecast accuracy with the interval score \citep{GneitingRaftery2007, BracherEtAl2021}. In particular, we consider quantiles at levels $\alpha \in \{0.1, 0.9\}$, which together form the central 80\% prediction interval. Such prediction intervals have been found to be a useful format for communicating probabilistic information to both expert and non-expert users \citep[see][and the references therein]{Raftery2016}. Specifically, let $l$ and $u$ denote the lower and upper bound of the prediction interval, and $y$ denote the realizing outcome. The interval score is then given by 
\begin{equation}
	\text{IS}(l,u,y) = (u-l) + 10 \times (l-y) \times \mathbf{1}(y < l) + 10 \times (y-u) \times \mathbf{1}(y > u),
	\label{intsc}
\end{equation}
with smaller scores being preferable. The score thus rewards short prediction intervals (with $u-l$ small) that nevertheless cover the realizing outcome $y$. The weight of ten on the penalty terms for not covering the outcome (as represented by the two indicator functions) ensures that the score is proper. That is, a forecaster minimizes their expected score by stating what they think is the true prediction interval.\footnote{The weight of ten for the penalty terms in Equation (\ref{intsc}) reflects the target interval coverage of $80 \%$. More generally, for the central $\kappa \%$ prediction interval, with $0 < \kappa < 100$, the weight is given by $200/(100-\kappa)$.} The interval score relates to the CRPS scoring rule mentioned earlier: While the interval score is proportional to the sum of two quantile scores \citep{Koenker1978,Gneiting2011} at levels $0.1$ and $0.9$, the CRPS is the unweighted integral over quantile scores at all levels $(0,1)$ \citep{GneitingRanjan2011}.

The interval score is a summary measure of forecast performance, reflecting both sharpness (i.e., the informativeness of the forecast) and calibration (i.e., the consistency between the forecast and the outcomes). For prediction intervals, sharpness is represented by the length of the intervals, and calibration is represented by the intervals' coverage rate. In our simulation and empirical analyses, we thus report these two measures in addition to the interval score. For forecast case $i = 1, \ldots, n$, denote the lower and upper bound of the prediction interval by $l_i$ and $u_i$, and the associated realization by $y_i$. The average length of the prediction intervals is then given by $\text{AL} = n^{-1} \sum_{i=1}^n (u_i-l_i)$. The coverage rate is given by $\text{CR} = n^{-1} \sum_{i=1}^n \mathbf{1}(l_i \le y_i \le u_i)$, i.e., the share of observations for which the prediction interval covers the corresponding realization. Assuming that $\text{CR} < 1$, the average interval score for an evaluation sample of size $n$ can be written as 
\begin{eqnarray}
	\frac{1}{n}\sum_{i=1}^n \text{IS}(l_i, u_i, y_i) = \text{AL} + 10 \times (1-\text{CR}) \times {\text{ASF}}_{\text{NC}},
\end{eqnarray}
where ${\text{ASF}}_{\text{NC}} = \frac{1}{n(1-\text{CR})} \sum_{i=1}^n \text{sf}_i,$ with
\begin{eqnarray*}
	\text{sf}_i &=& (l_i-y_i) \mathbf{1}(y_i < l_i) + (y_i-u_i) \mathbf{1}(y_i > u_i),
\end{eqnarray*}
i.e. ${\text{ASF}}_{\text{NC}}$ denotes the average shortfall (distance between observation and nearest end of prediction interval) in case of non-coverage. 

\subsection{Sample split for model estimation}

A model's forecast performance should generally be assessed on data points that were not used for model fitting. In time series contexts, it is common to use rolling or expanding samples of data for estimation.\footnote{In a rolling sample, a one-step forecast for period $t$ is based on a training sample ranging from period $t-R$ to $t-1$, where $R$ is the length of the rolling sample. In an expanding sample, the traning sample ranges from periods $1$ to $t-1$, thus expanding over time.} Such model validation strategies are attractive if the time series of interest is sufficiently long. For short time series, these strategies are less attractive: Long (rolling or expanding) estimation samples mean that only few observations are left for model evaluation, yielding low power in model comparisons. Short estimation samples, on the other hand, may yield erratic model fits and unstable results. We seek to avoid these drawbacks, but nevertheless achieve a clear separation between training and test data. We thus use a cross-validation approach where predictions referring to year $t$ (i.e., referring to forecast errors $e_{t,h_j}$ for various horizons $h_j$) are based upon an estimation sample that comprises all years other than $t$. \cite{BergmeirEtAl2018} argue in favor of similar cross-validation strategies in the context of autoregressive time series models. 

\section{Simulation results}\label{sec:sim}

We next conduct a simulation study in order to assess the finite-sample properties of our methodology, including the models from Section \ref{sec:models} and the evaluation techniques from Section \ref{sec:eval}. To this end, we simulate data from the model presented in Section \ref{sec:ar1model}, as well as optimal mean forecasts $X_{t,h}$ (see Section A.4 of the online supplement) and associated forecast errors $e_{t,h}$ for various horizons $h$. We then construct a random sample $\{e_{t_i, h_i}\}_{i=1}^n$ of forecast errors, where $t_i$ is drawn uniformly from $\{1, 2, \ldots, T_{\text{max}}\}$, and $h_i$ is drawn uniformly from $\{0, 1, \ldots, 104\},$ corresponding to a maximal forecast horizon of two years. This setup aims to mimic the heterogeneous and overlapping nature of the sample of forecast errors as described in Section \ref{sec:sample}. We set $(n, T_\text{max})$ to either $(500, 30),$  $(1000, 30)$ or $(500, 60)$. Compared to the first sample type, the second type contains more observations from the same time period, whereas the third type contains the same number of observations from a longer time period. We further use the parameter values $\rho = 0.3, \sigma^2_\varepsilon = 0.09$ and $\sigma^2_\eta = 0.003$ that were also used in Figure \ref{fig:rmse}. Finally, we use a burn-in period of $30$ years in order to remove the impact of the prior parameters used to initialize the Kalman filter. 

In addition to the error postprocessing methods described in Section \ref{sec:models}, we consider the true forecast error distribution that is implied by the structural model from which the data is simulated. Importantly, this distribution is not realistically available in practice, as it requires knowledge of the model's functional form and its true parameter values. The true distribution hence defines an unattainable gold standard that allows to set the other methods' performance in perspective.

\begin{table}
	\centering
	\textit{Setup 1: $n = 500, T = 30$} \\[.3cm]
	\begin{tabular}{lccc}
		Model & Coverage & PI length & Interval Score \\ \toprule
		Gaussian & 77.85\% & 4.03 & 5.71 \\
		Decomposition & 78.06\% & 4.04 & 5.72 \\
		Flexible & 76.10\% & 3.94 & 6.01 \\
		Combination & 78.38\% & 4.00 & 5.72 \\ \midrule
		True & 80.00\% & 4.07 & 5.58 \\\bottomrule
	\end{tabular}\\[.5cm]
	\textit{Setup 2: $n = 1\,000, T = 30$}\\[.3cm]
	\begin{tabular}{lccc}
		Model & Coverage & PI length & Interval Score \\ \toprule
		Gaussian & 78.05\% & 4.02 & 5.68 \\
		Decomposition & 78.58\% & 4.04 & 5.68 \\
		Flexible & 76.63\% & 3.94 & 5.93 \\
		Combination & 78.65\% & 4.00 & 5.69 \\ \midrule
		True & 80.08\% & 4.07 & 5.56 \\\bottomrule
	\end{tabular}\\[.5cm]
	\textit{Setup 3: $n = 500, T = 60$} \\[.3cm]
	\begin{tabular}{lccc}
		Model & Coverage & PI length & Interval Score \\ \toprule
		Gaussian & 78.55\% & 4.05 & 5.65 \\
		Decomposition & 78.36\% & 4.05 & 5.68 \\
		Flexible & 76.79\% & 3.94 & 5.89 \\
		Combination & 78.72\% & 4.01 & 5.67 \\ \midrule
		True & 79.96\% & 4.07 & 5.58 \\\bottomrule
		&&&\\ \vspace{-.6cm}
	\end{tabular}
	\caption{Simulation results on forecast performance. `Coverage' denotes the share of realizing observations that fall into a model's central 80\% prediction intervals. `PI length' is the average length of a model's prediction intervals. `Interval Score' denotes the score in Equation (\ref{intsc}), with smaller values being preferable. Results are based on $2\,000$ Monte Carlo iterations, using cross-validation for each simulated sample. See Section \ref{sec:eval} for details on the evaluation methodology. Results are averaged across all forecast horizons.\label{fig:mc}}
\end{table}

Table \ref{fig:mc} displays the simulation results. The findings on the interval score (rightmost column of the table) can be summarized as follows. First, and as expected, the true model outperforms the three methods that are estimated based on samples of forecast errors.  Second, the prediction methods' performance tends to improve as the sample becomes more informative (either by covering a longer time span or by covering more forecast errors from the same time span). The added value of training data is especially large for the most flexible method. Note that the expected performance of the true model is the same across all sample types, and any observed differences are within the range of Monte Carlo error.\footnote{Using two-sample $t$-tests, performance differences of the true model across any pair of sample types are not statistically significant at conventional levels.} Third, the Gaussian, decomposition and combination methods outperform the flexible method for all three sample types. The good performance of the decomposition method can be explained by its use of constraints (in particular, symmetry and a zero mean of forecast errors) that are satisfied by the true model, so that they serve to regularize the estimator. While the assumptions of the Gaussian method (in particular, the specification of the standard deviation $\sigma_h$) are not exactly satisfied by the true model, the degree of misspecification seems minor, so that the Gaussian method also benefits from regularization. 

The results in Table \ref{fig:mc} further indicate that the Gaussian, decomposition and combination methods mostly reach good coverage rates of 78\% or more, as well as prediction intervals of similar length. The flexible method's coverage rates are slightly worse, and its prediction intervals are somewhat shorter. 

In the present simulation setup, the methods we consider thus yield plausible results, despite the challenges posed by small sample sizes and overlapping data.

\section{Empirical results}\label{sec:emp}

\subsection{Data}\label{sec:data}

We present results for two empirical data sets consisting of fixed-event point forecasts and associated realizations. 

First, we consider a data set covering forecasts of German real GDP growth, as made by ten institutions: The Bundesbank, European Central Bank (ECB), European Comission (EC), International Monetary Fund (IMF), Organisation for Economic Co-operation and Development (OECD), as well as three German research institutes (DIW Berlin, ifo Munich, and IWH Halle) and two committees that play a prominent role in the policy debate (the German council of economic experts, and the Gemeinschaftsdiagnose, a joint forecast made by several research institutes). We obtain all forecasts and the corresponding first-release outcome data by the IWH Halle's forecasting dashboard \citep{dashboard,HeinischEtAl2023}, a recent initiative that makes a rich archive of German economic forecasts openly available. The data indicate the origin date (day) and target date (year) of each forecast, so that precise information on the forecast horizon is available. We express the forecast horizon in weeks, noting that all methods we consider can accommodate non-integer forecast horizons (such as $h = 25/7$ weeks representing $25$ days). Various studies \citep[e.g.][]{DoepkeFritsche2006,KoehlerDoepke2022} consider the properties of such fixed-event point forecasts for Germany. \cite{Foltas2022} analyze whether the distribution of forecast errors can be predicted by means of regressor variables. However, their analysis seeks to test (a particular notion of) forecast efficiency, as opposed to constructing forecast distributions. Furthermore, they treat each forecast horizon separately, while interpolating across horizons is the key methodological feature of our approach. 

Importantly, we treat the forecasting institutions as exchangeable, that is, we do not attempt to model the forecast error as a function of the institution that made the forecast. This choice is motivated by empirical results from the forecast combination literature which suggest that treating economic forecasters as exchangeable often performs well in terms of the bias-variance trade-off \citep[e.g.][]{GenreEtAl2013,ClaeskensEtAl2016}. We expect similar findings to hold in the present case, especially in view of the rather small sample size. In the cases where two or more institutions make a forecast on the same day (11.2\% of observations), we keep all forecasts. This means that we seek to predict the error made by a randomly drawn forecasting institution, as opposed to the forecast error made by an ensemble. This approach seems most practically relevant in the present context, in that most days feature at most one forecast.

Second, we consider US data from the Survey of Professional Forecasters \citep{Croushore2019} covering GDP and inflation forecasts from 1981 to 2021. The SPF's point forecasts are widely considered a hard-to-beat benchmark for even sophisticated statistical forecasting models \citep[e.g.][]{FaustWright2013}. Here we use fixed-event point forecasts that are available on a quarterly basis, referring to the present and next year. Since the forecasts are made roughly in the middle of a quarter (with each quarter corresponding to $52/4 = 13$ weeks), the forecast horizons satisfy $h \in \{6.5, 19.5, \ldots, 97.5\}$. A specific pair of two horizons is available each quarter. For example, in the first quarter of a year, the current-year forecast corresponds to $h = 45.5$ weeks, and the next-year forecast corresponds to $h = 97.5$ weeks. Furthermore, we consider the SPF forecast distributions that cover participants' subjective probabilities for various ranges of the outcome variable. We use the average survey response for both point and probabilistic forecasts from the SPF. We drop data from 1985:Q1, 1986:Q1 and 1990:Q1 due to possible technical errors in the associated survey rounds \citep{SPF_docu}. At each forecast date, we approximate the probabilistic forecasts by a continuous distribution, following \cite{EngelbergEtAl2009} and using the implementation of \citet[Appendix A]{KruegerPavlova2022}. We use first-release data provided by the Federal Reserve Bank of Philadelphia to compute the actual GDP growth and inflation outcomes.

Table \ref{tab:datasets} provides summary information about the data sets. Given the lack of easily available probabilistic forecasts for German GDP, the first data set is particularly interesting from a policy perspective. The US SPF data provide a useful testbed to assess the performance of our statistical models since they cover a longer time period with more variation in macroeconomic outcomes. Furthermore, the SPF's probabilistic forecasts provide a natural benchmark for our postprocessing methods. 

\begin{table}
	\centering
	\begin{tabular}{lcccrrrrr}
		\toprule
		&  \multicolumn{2}{c}{Target year} & Sample size & \multicolumn{5}{c}{Forecast horizon $h$ (weeks)} \\
		Data set & Min. & Max. &  & Min. & 25\% & 50\% & 75\% & Max.\\
		\midrule
		German GDP & 1991 & 2022 & 1307 & 0.0 & 23.7 & 49.4 & 74.3 & 104.0\\
		US GDP & 1981 & 2022 & 320 & 6.5 & 19.5 & 45.5 & 71.5 & 97.5\\
		US Inflation & 1981 & 2022 & 320 & 6.5 & 19.5 & 45.5 & 71.5 & 97.5\\	
		\bottomrule
	\end{tabular}
	\caption{Summary information on forecast/realization data sets considered in this paper. German GDP data are from the IWH Halle Forecasting dashboard \citep{HeinischEtAl2023}. US GDP and inflation data are from the Survey of Professional Forecasters. For `Sample size', one forecast/observation pair counts as one observation. \label{tab:datasets}}
\end{table}

\begin{table}
	\centering
	\textit{German GDP (1991-2022, $n = 1307$)} \\[.3cm]
	\begin{tabular}{lccc}
		Model & Coverage & PI length & Interval Score \\ \toprule
		Gaussian & 79.11\% & 2.71 & 5.81 \\
		Decomposition & 79.27\% & 2.85 & 5.92 \\
		Flexible & 78.35\% & 2.97 & 6.35 \\
		Combination & 81.48\% & 2.84 & 5.94 \\ \bottomrule
	\end{tabular}\\[.5cm]
	\textit{US GDP (1981-2022, $n = 320$)}\\[.3cm]
	\begin{tabular}{lccc}
		Model & Coverage & PI length & Interval Score \\ \toprule
		Gaussian & 76.56\% & 2.30 & 4.11 \\
		Decomposition & 79.06\% & 2.38 & 4.06 \\
		Flexible & 75.31\% & 2.08 & 4.10 \\
		Combination & 76.88\% & 2.26 & 4.07 \\\midrule
		SPF Histogram & 85.94\% & 2.97 & 4.48 \\ \bottomrule
	\end{tabular}\\[.5cm]
	\textit{US Inflation (1981-2022, $n = 320$)} \\[.3cm]
	\begin{tabular}{lccc}
		Model & Coverage & PI length & Interval Score \\ \toprule
		Gaussian & 78.44\% & 1.30 & 2.65 \\
		Decomposition & 78.75\% & 1.33 & 2.67 \\
		Flexible & 72.19\% & 1.13 & 2.77 \\
		Combination & 78.44\% & 1.25 & 2.66 \\ \midrule
		SPF Histogram & 85.94\% & 2.25 & 3.35 \\\bottomrule
		&&&\\ \vspace{-.6cm}
	\end{tabular}
	\caption{Empirical results on forecast performance. `Coverage' denotes the share of realizing observations that fall into a model's central 80\% prediction intervals. `PI length' is the average length of a model's prediction intervals. `Interval Score' denotes the score in Equation (\ref{intsc}), with smaller values being preferable. All results are based on cross-validation as described in Section \ref{sec:eval}. Results are averaged across all empirical cases (referring to various forecast horizons). \label{tab:scores}}
\end{table}

\subsection{Main results}

Table \ref{tab:scores} presents our main empirical results, which are pooled across all forecast horizons. For the German data, all forecasting methods attain coverage rates close to the nominal level of $80 \%$. The Gaussian method produces the shortest prediction intervals, while the flexible method produces the widest ones. The interval score is highest (i.e., worst) for the flexible method. For the US data (GDP and inflation), the Gaussian, decomposition, and combination methods again attain coverage rates close to 80\%, whereas the flexible method's rater are lower. On the other hand, the flexible method produces shorter prediction intervals, so that its overall performance (in terms of the interval score) is similar to the other three methods. 

For the Gaussian method, we also considered a restricted variant with mean $\mu = 0$, i.e., assuming unbiased mean forecasts. Its performance (reported in Table S2 in the online supplement) is very similar to that of the unrestricted model. 

In order to investigate whether the methods' performance varies across forecast horizons $h$, Figure S2 in the online supplement plots their coverage rate against $h$. As shown by the figure, the Gaussian, decomposition and combination methods attain good coverage rates (close to 80\%) across all horizons. The flexible method's coverage rates tend to be slightly too low for the longest horizons. 

We next compare the error-based forecasting methods to the SPF's histogram-type forecasts that are available for the two US data sets. As shown in Table \ref{tab:scores}, the histograms' coverage rate exceeds its nominal level of 80\%, which by itself is desirable. However, this coverage rate comes at the expense of very wide prediction intervals. This applies especially to inflation, where the average length of the histogram intervals exceeds the average length of the other methods' intervals by about 75\%. Furthermore, the histograms attain markedly higher interval scores than the other methods. 

\begin{figure}
	\begin{tabular}{cc}
		\textit{US GDP} & \textit{US Inflation} \\ 
		\includegraphics[width=.5\textwidth]{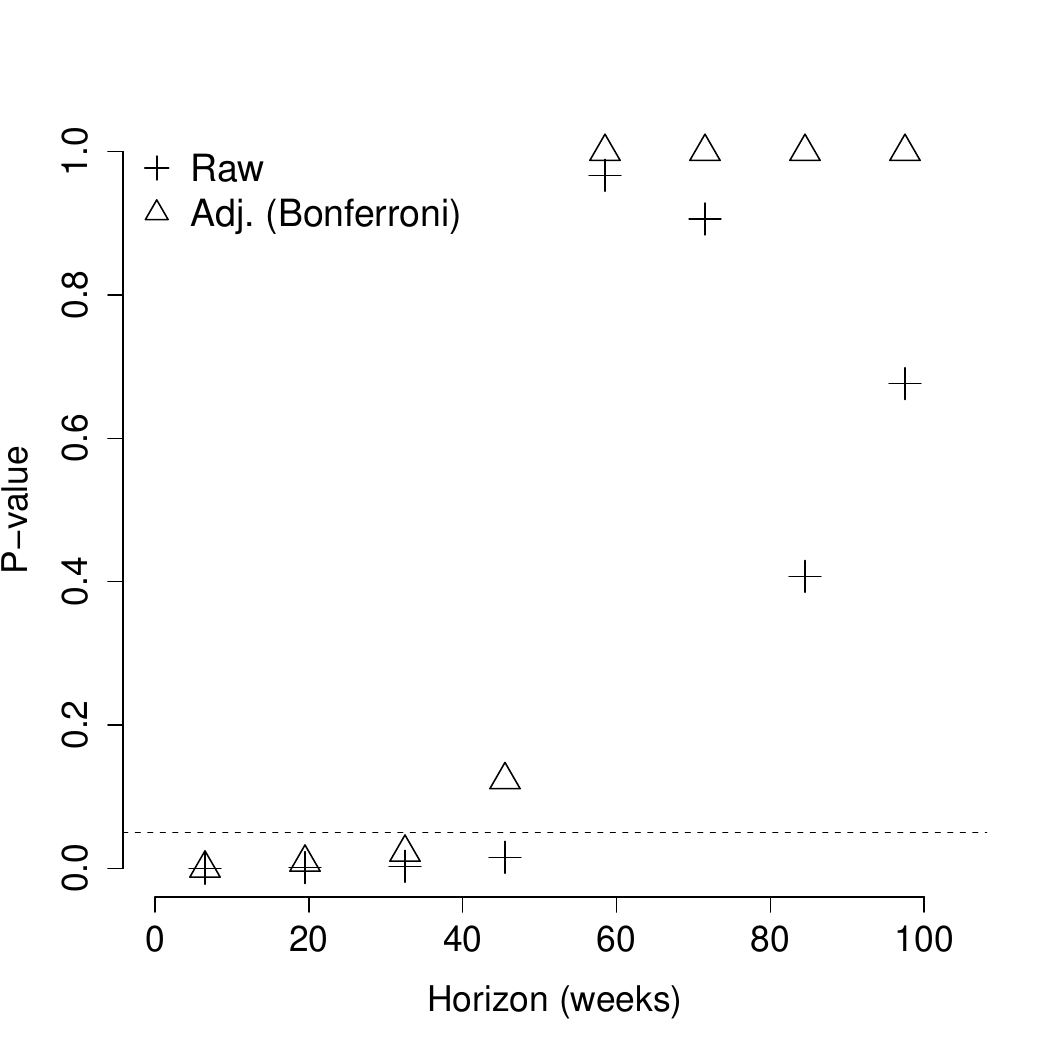} & 
		\includegraphics[width=.5\textwidth]{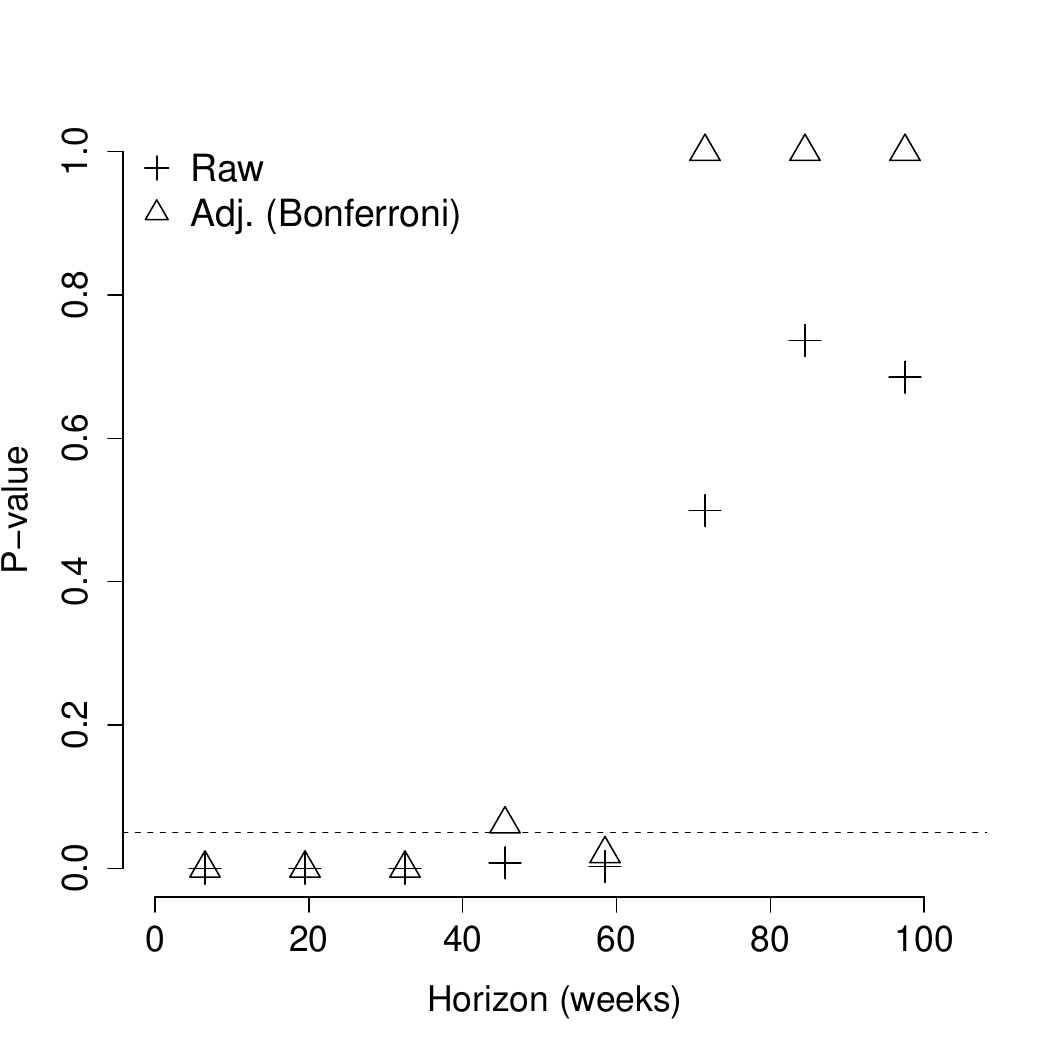} \\
	\end{tabular}
	\caption{$P$-values of Diebold-Mariano type tests. Under the null hypothesis, the combination method (based on error postprocessing) and the SPF histogram have equal predictive ability. Two-sided tests are conducted separately across eight forecast horizons. Crosses: Raw (un-adjusted) $p$-values; triangles: $p$-values adjusted for multiple testing via the Bonferroni method. Dashed horizontal line indicates a significance level of five percent. All rejections at the 5\% level are in favor of the combination method.\label{tab:dm}}
\end{figure}

In order to investigate whether these performance differences are statistically significant, we conduct \cite{DieboldMariano1995} type tests, using the combination method as a natural representative of the postprocessing methods. As noted in Section \ref{sec:sample}, the dependence structure of forecast errors is potentially complex when pooling the results across forecast horizons. Similar complexity is to be expected for \citeauthor{DieboldMariano1995} type tests, which are based on functions of forecasts and realizations (specifically, interval score differences). In order to reduce complexity, and arrive at the standard time series setup considered by \citeauthor{DieboldMariano1995}, we compare the combination to the SPF histograms separately for each forecast horizon. This choice comes at the cost of a smaller evaluation sample of a single forecast/observation pair per year for each horizon. Due to some variation in data availability across horizons, the size of the evaluation samples then ranges from $36$ to $42$. The test statistics are based on autocorrelation-consistent standard errors as implemented in the function \textsf{NeweyWest} of the R package \textsf{sandwich} \citep{Zeileis2004,ZeileisEtAl2020}. Figure \ref{tab:dm} summarizes the results. For GDP, the combination significantly outperforms the SPF at the three shortest horizons, using a 5\% significance level and a conservative (Bonferroni) $p$-value adjustment for multiple testing. For inflation, we similarly observe significant outperformance at four of the five shortest horizons. For the other horizons, the performance difference between the combination and the SPF is insignificant. In Section B.2 of the online supplement, we provide additional discussion and results on \cite{DieboldMariano1995} type testing in the current setup, including three additional implementation variants for computing the test statistic. As described in the supplement, these variants differ in their treatment of potential autocorrelation in the time series of interval score differences. While some of the variants yield considerably higher $p$-values for GDP at the four shortest horizons, the results for inflation are similar to the ones in Figure \ref{tab:dm}.

\begin{figure}
	\centering
	\begin{tabular}{cc}
		\textit{US GDP} & \textit{US Inflation} \\[-.5cm]
		\includegraphics[width=.5\textwidth]{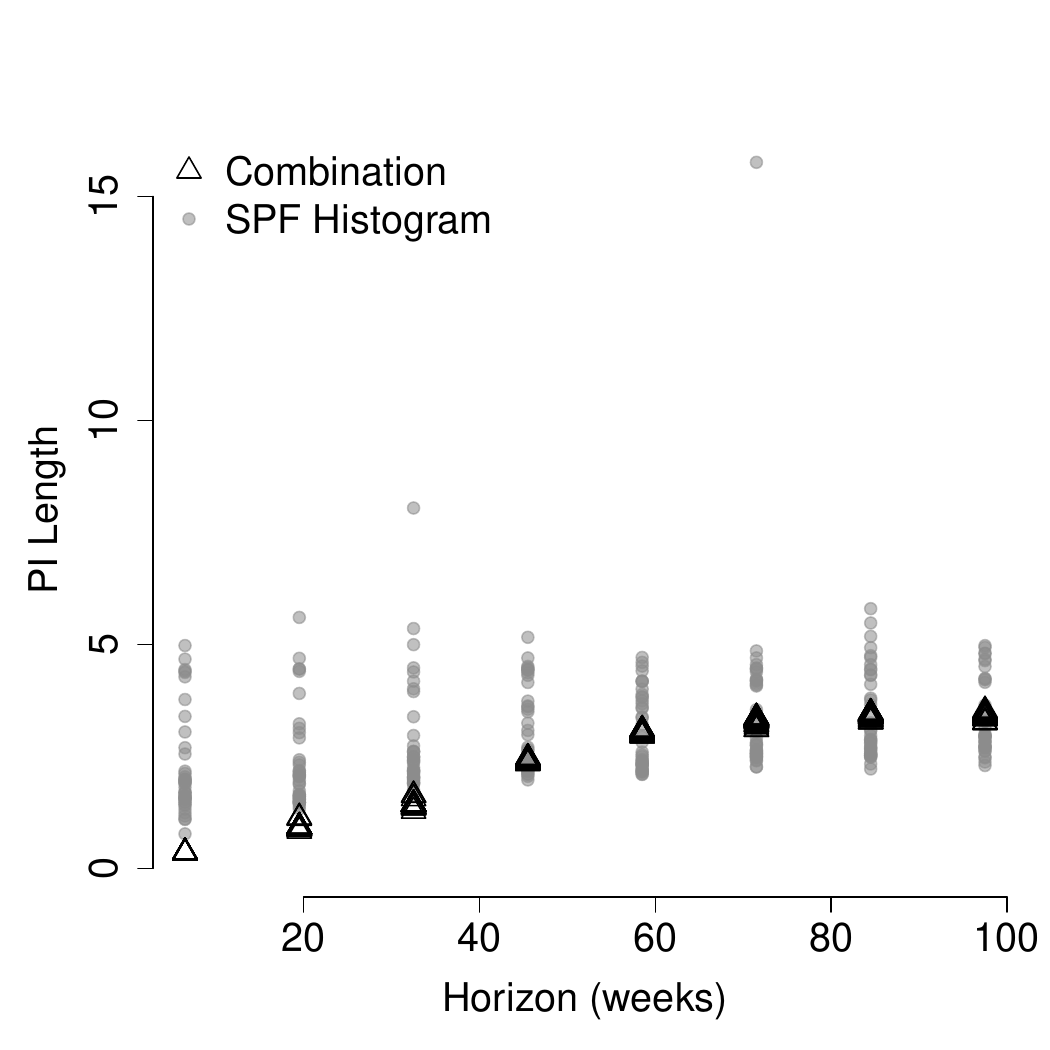} & 
		\includegraphics[width=.5\textwidth]{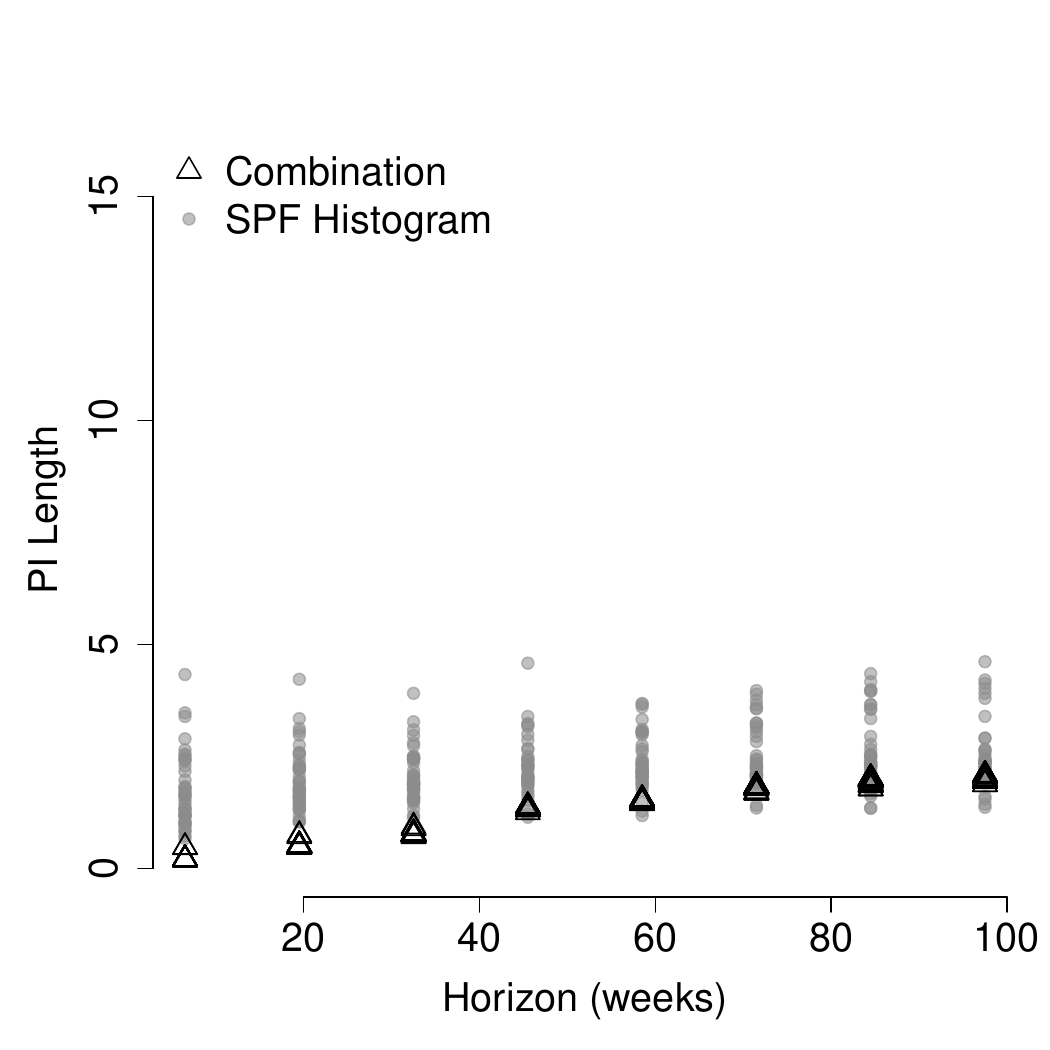}
	\end{tabular}
	\caption{Length of prediction intervals plotted against forecast horizon. Each dot or triangle represents one forecast case. The outlier in the left panel (PI length $> 15$) corresponds to the SPF's GDP forecast for 2021 made in August 2020.\label{fig:pilength}}
\end{figure}

Figure \ref{fig:pilength} further shows the length of the combination and SPF methods' prediction intervals separately for each horizon.\footnote{Each dot or triangle in the figure represents a single forecast case. For the combination method (represented by triangles), the intervals are quite similar in length across forecast cases; this is because of highly overlapping training samples resulting from the employed cross-validation procedure. The variation across forecast cases is hence dwarfed by the amount of variation in the SPF, and is often invisible in the figure.} The SPF's intervals are clearly wider than the combination's, especially at short forecast horizons. Findings from the literature on combining forecast distributions \citep[e.g.][]{GneitingRanjan2013,Lichtendahl2013} suggest that the width of the SPF's prediction intervals may partly be driven by our use of \textit{linear} combination, in that we consider the \textit{mean} of the SPF participants' probability forecasts.\footnote{Note, however, that we fit a continuous distribution to these mean probability forecasts in order to compute quantiles. This extra fitting step is not covered by the theoretical literature on combining forecast distributions.} It is thus natural to ask whether nonlinear combination methods are more successful in the context of the SPF. To this end, we consider two quantile-based combination approaches: First, the simple average of the individual participants' quantile forecasts, as introduced in Section \ref{sec:comb} in the context of combining postprocessing methods. Second, the median of the individual quantile forecasts \citep[see e.g.][]{BracherEtAl2021}. Both approaches require us to first fit a parametric distribution to each SPF participant's forecast distribution, for which we again follow \citet[Appendix A]{KruegerPavlova2022}. As shown in Table S2 in the online supplement, the quantile-based combination methods indeed yield shorter prediction intervals. However, this comes at the cost of a  lower coverage rate, so that they attain similar interval scores as the linear SPF combination, and thus perform worse than the postprocessing methods. 

The findings discussed in the previous two paragraphs are closely in line with \cite{Clements2010,Clements2014} who documents that the SPF's fixed-event forecast distributions are implausibly wide at short horizons, perhaps reflecting incoherent updating behavior on the part of forecasters. On the whole, our results thus indicate that even fairly simple models based on past forecast errors are more accurate than the forecasters' own assessment of uncertainty.

\subsection{Illustrations}

\begin{figure}
	\centering
	\begin{tabular}{c}
		\includegraphics[width=.6 \textwidth]{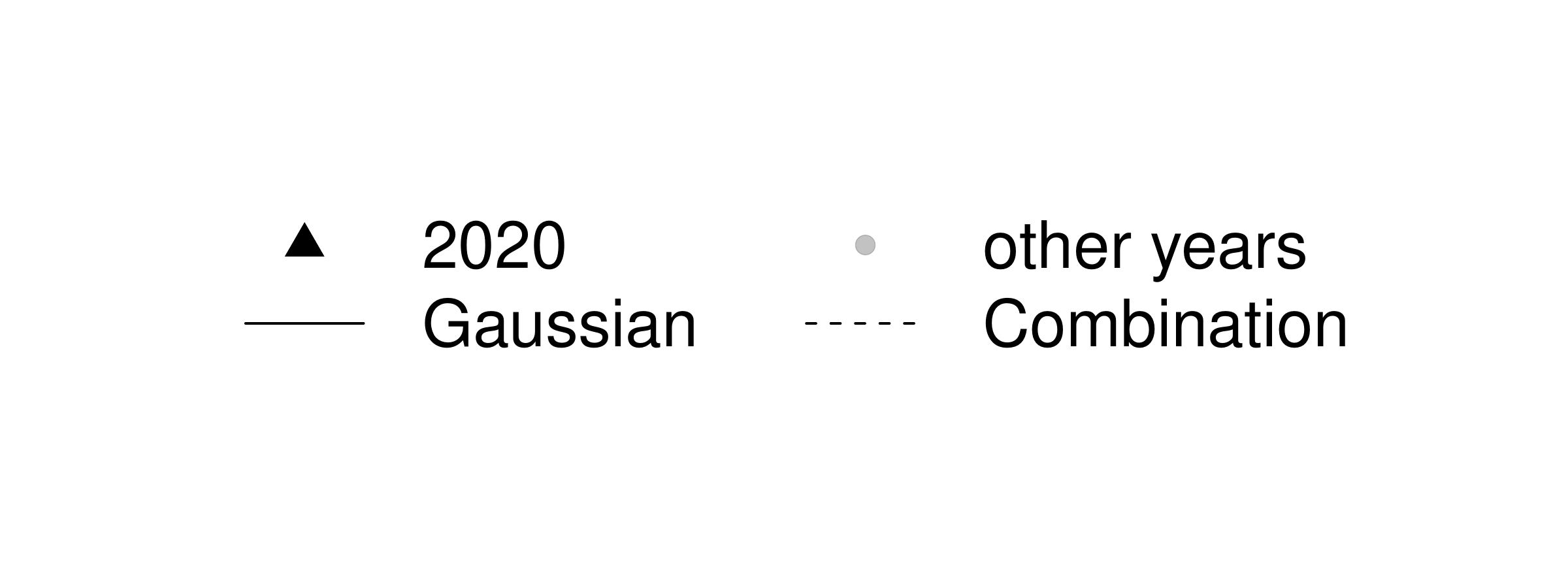}\\[-.5cm]
		\textit{German GDP} \\[-.7cm]
		\includegraphics[width=\textwidth]{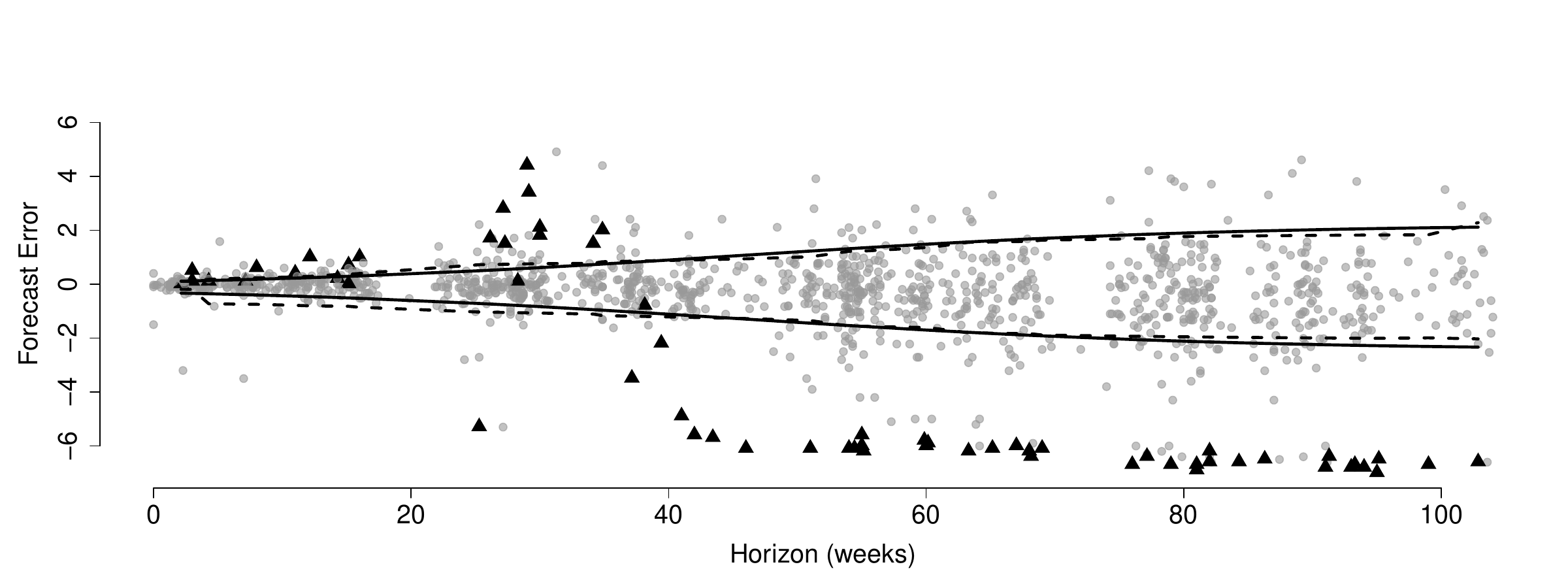}\\
		\\[.2cm]
		\textit{US GDP} \\[-.7cm]
		\includegraphics[width=\textwidth]{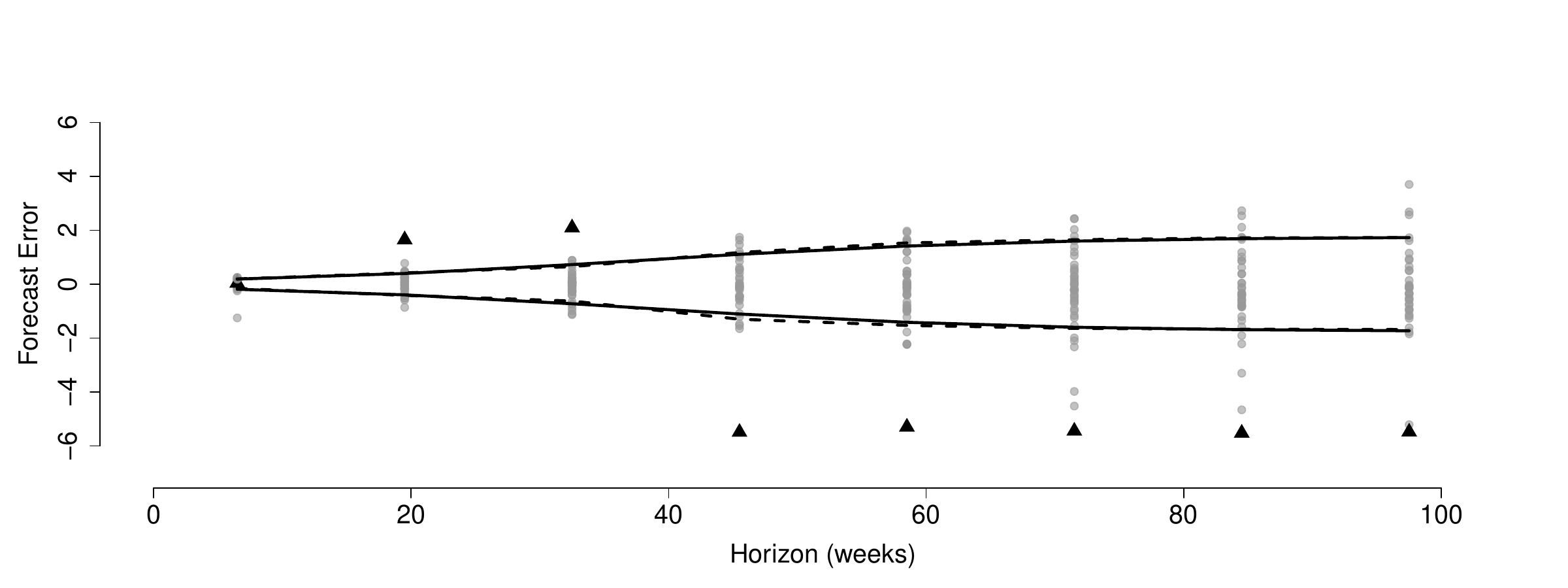}\\
		\\[.2cm]
		\textit{US Inflation} \\[-.7cm]
		\includegraphics[width=\textwidth]{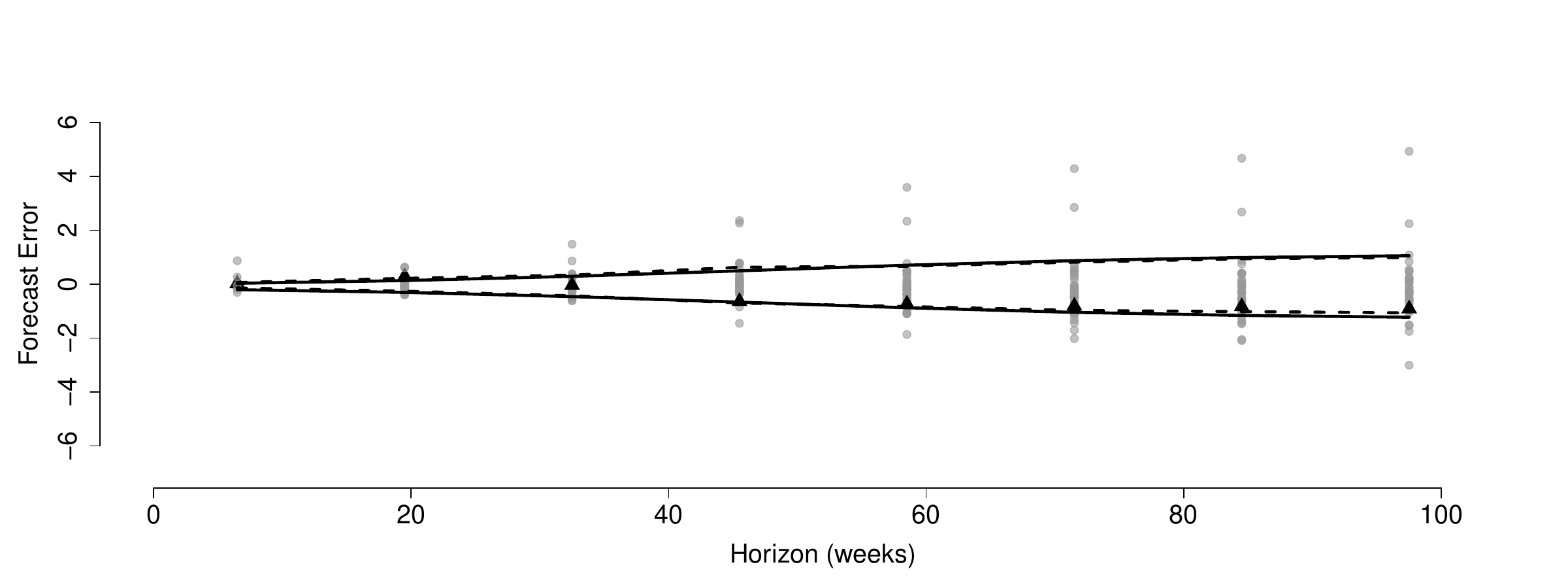}
	\end{tabular}
	\caption{Forecast errors plotted against forecast horizon. Curves indicate estimated 10\% and 90\% quantiles using two methods (solid curves: Gaussian, dashed curves: Combination). Model estimates are from cross-validation iteration that omits 2020 data. Black triangles represent 2020 realizations, grey dots represent realizations from other years.\label{fig:2020}}
\end{figure}

Figure \ref{fig:2020} illustrates the Gaussian and combination methods for making prediction intervals, for German GDP (top panel), US GDP (middle) and US inflation (bottom). The figure refers to the cross-validation run that excludes the 2020 data from model fitting. Thus, the figure contains both test-sample observations (for 2020, represented by black triangles) and training-sample observations (other years, represented by grey dots). Partly by construction, the clear majority of historical forecast errors is within the methods' prediction intervals.\footnote{For the training-sample observations, close-to-nominal coverage is encouraged by the criteria used for fitting the models. Hence the models' good coverage properties for these observations arises `partly by construction'. By contrast, note that the coverage rates reported in Table \ref{tab:scores} are computed from test-sample observations exclusively, and hence do not arise by construction.} For the Gaussian method (solid line), the intervals are symmetric around zero. For the combination method (dashed line), the intervals are somewhat asymmetric, especially for German GDP. The figure also shows that the 2020 forecast errors for German and US GDP are very large by the standards of the data set. This applies in particular to the negative forecast errors at horizon $41$ or larger (corresponding to forecasts made before mid March 2020), and can be attributed to the effects of the Covid-19 pandemic. By contrast, some of the 2020 forecast errors at horizons 20-30 are large and positive, indicating that forecasts made around mid-2020 were too pessimistic. It seems unsurprising that the methods' prediction intervals -- which are designed to attain an 80\% coverage level -- do not capture forecast errors in an excessively turbulent setup like 2020 GDP. For US inflation (bottom panel), 2020 forecast errors are mostly covered by the methods' prediction intervals.

To illustrate the quantitative interpretation of our results, consider the decomposition method as applied to German GDP, and a hypothetical forecast made in mid-September. The 80\% prediction intervals for the current-year forecast (corresponding to horizon $15$) are $0.68$ percentage points wide, and a point forecast of $m \%$ GDP growth translates into a prediction interval of $[(m-0.34)\%, (m+0.34)\%]$. The prediction intervals for the next-year forecast (corresponding to horizon $67$) are more than five times as wide, and are given by $[(m-2.01)\%, (m+2.01)\%]$. In forecast-related press statement or media reports, point forecasts for the current and next are often mentioned next to each other. Our numerical example highlights that this practice is possibly misleading, in that forecast uncertainty may differ substantially across the two horizons.

\section{Discussion}\label{sec:disc}

This paper argues that economic fixed-event forecasts should be accompanied by a numerical measure of uncertainty, and proposes methods for computing such a measure. We conclude by discussing relations to the pertinent literature, as well as practical implications.

\textit{Transforming fixed-event forecasts.} A number of studies consider transforming fixed-event forecasts into fixed-horizon forecasts, on the grounds that the latter are more convenient and more flexible from a statistical perspective. \cite{KnueppelVladu2016} consider transforming fixed-event point forecasts, whereas \cite{GanicsEtAl2020} seek to transform the SPF's fixed-event forecast distributions. \cite{ClarkEtAl2022} use entropic tilting to incorporate fixed-event information from the SPF (both point forecasts and distributions) into a statistical model that can generate flexible types of forecasts. In contrast to these studies, we estimate the uncertainty of fixed-event point forecasts as a function of their horizon $h$.

\textit{Pooling versus not pooling of forecast error data.} Our methods are based on pooling past forecast errors across horizons $h$, and fitting the distribution of forecast errors as a function of $h$. This approach is motivated by a small sample of forecast errors for each specific horizon $h$, especially for the German data set we consider. In more data-rich situations, using entirely separate statistical models for each forecast horizon may be preferable. \cite{DieboldGoebel2022} follow this alternative approach for constructing fixed-event predictions of arctic sea ice extent.  Furthermore, we assume that the distribution of forecast errors depends on $h$ alone, which could be relaxed in more data-rich situations. For example, allowing for heterogeneity across target years (by means of random effects type models, say) is conceivable.

\textit{Communication of forecast uncertainty.} Many economists agree that forecast uncertainty should be measured and communicated. According to \cite{ReifschneiderTulip2019}, for example, the fact that `[..] prediction errors -- even on occasion quite large ones -- are a normal part of the process [..]' should be communicated to the public in order to enhance the credibility of future forecasts. Perhaps most prominently, many central banks publish forecast distributions (`fan charts') pioneered by the Bank of England in the 1990s \citep[see e.g.][]{GalbraithNorden2012}. That said, most economic point forecasts are not accompanied by a numerical measure of uncertainty. Instead, media reports and publications by forecasting institutions often contain a verbal disclaimer that mentions forecast uncertainty. For example, the European Commission's Summer 2022 forecast lists various `risks to the outlook' \citep[Section 1.6]{EuropeanCommission2022}, representing possible economic sources of forecast error. In our view, verbal statements are not a satisfactory alternative to a numerical measure in the present context. Among other problems,  words such as `likely' often mean different things to different people, and are hard to falsify ex-post \citep[see e.g.][]{DhamiMandel2022}. Communicating forecast uncertainty in a practical yet precise way thus remains an important challenge in economic policy. Prediction intervals are an attractive format for doing so since they are reasonably simple (consisting of two numbers only), can be represented in graphical form, and are easy to evaluate ex-post \citep{Raftery2016}.

\setcounter{figure}{0}
\setcounter{table}{0}

\renewcommand{\thefigure}{S\arabic{figure}}	
\renewcommand{\thetable}{S\arabic{table}}	

\clearpage

\begin{appendix}
	
\noindent{\textbf{\huge Online Supplement}}\\
	
\noindent Section \ref{sec:sec3} of the online supplement provides details on the model from Section 3 of the main paper, whereas Section \ref{sec:empos} provides additional empirical analysis and results.

	\section{Details on the autoregressive model}
	
	\label{sec:sec3}
	
	\subsection{Comparison to \cite{PattonTimmermann2011}} \label{app:pt}
	Our model in Section 3 of the main paper differs from the one by \citeauthor{Patton2012} in three respects. First, we use weekly (rather than monthly) high-frequency observations. Second, while \citeauthor{PattonTimmermann2011} consider both independent and serially correlated measurement error, we focus on the independent case which seems sufficient to capture the stylized facts of interest. Finally, in our setup the coefficients $\gamma_j$ (defined below Equation 3 in the main paper) are either strictly increasing or strictly decreasing in $j$, whereas \citeauthor{PattonTimmermann2011}'s weight function for GDP features flat segments (see their Figure B.1). This difference arises because \citeauthor{PattonTimmermann2011} assume that the quarterly GDP index is given by the index of the quarter's last month (e.g., March for the year's first quarter). By contrast, we represent a quarter's index by the average of all weeks within the quarter (e.g., weeks 1-13 for the first quarter). We prefer our functional form since it is simpler while, in our view, being at least as plausible. In particular, averaging the weekly levels within a quarter is consistent with averaging the quarterly levels within a year (which is done by both \citeauthor{PattonTimmermann2011}'s and our approach). 
	
	\subsection{Derivation of aggregation weights}\label{app:w}
	
	Here we derive the triangular weighting function that we use to 
	approximate the annual-average-over-annual-average growth rate based on (hypothetical) weekly data. While this weighting function and its excellent approximation quality are well known \citep[see e.g.][and the references therein]{PattonTimmermann2011, HepenstrickBlunier2022}, our derivation is simpler than the ones we are aware of, and we thus include it for ease of reference.\\
	
	Let $\text{GDP}_w^*$ denote the hypothetical level of GDP in week $w$. The predictand of interest is the growth rate in the annual average of ¸GDP. We next show that this growth rate is well approximated by a weighted sum of weekly logarithmic growth rates. For ease of presentation, we consider the percent growth rate from year $t = 1$ (weeks $w=1, 2, \ldots, 52$) to year $t = 2$ (weeks $w =53, 54,\ldots, 104$), which we denote by $g_2$. We obtain
	
	\begin{eqnarray}
		g_2/100&=&\frac{\frac{1}{52}\sum_{w=53}^{104}\text{GDP}_w^*}{\frac{1}{52}\sum_{w=1}^{52}\text{GDP}_w^*}-1\\
		&\approx& \log\left(\frac{1}{52}\sum_{w=53}^{104}\text{GDP}_w^*\right) - \log\left(\frac{1}{52}\sum_{w=1}^{52} \text{GDP}_w^*\right)\label{approx1} \\
		&\approx & \log \left(\prod_{w=53}^{104}\text{GDP}_w^*\right)^{1/52} - \log \left(\prod_{w=1}^{52}\text{GDP}_w^*\right)^{1/52} \label{approx2}\\
		&=& \frac{1}{52}\sum_{w=53}^{104} \left(\log \text{GDP}_w^* - \log \text{GDP}_{w-52}^*\right)\label{secondtolast}\\
		&=& \sum_{w=2}^{104}\frac{52-|53-w|}{52}~\underbrace{\left(\log \text{GDP}_w^* - \log \text{GDP}_{w-1}^*\right)}_{\equiv Y_w^*/100},\\
		&=& \sum_{j=1}^{103}\underbrace{\left(1-\frac{|52-j|}{52}\right)}_{\equiv \gamma_j}Y^*_{105-j}/100,\label{last}
	\end{eqnarray}
	where $Y_w^*$ is the logarithmic GDP growth rate from week $w-1$ to $w$ (in percent), and $\gamma_j \in [1/52, 1]$ denotes a triangle-type weighting function as described in the main text. Note that Equation (\ref{last}) corresponds to Equation (3) from the main paper in the case $t=2$.
	
	The first approximation (Equation \ref{approx1}) replaces exact growth rates by logarithmic growth rates, based on the common first-order Taylor expansion $\log(1+z) \approx z$ for $z \approx 0$. The second approximation (Equation \ref{approx2}) replaces the arithmetic mean of the annual GDP level by the geometric mean. Note that both approximations are exactly correct if and only if GDP is constant over time, i.e. if $\text{GDP}_1^* = \text{GDP}_2^* = \ldots = \text{GDP}_{104}^*$. In this case, the annual growth rate $g_2$ is exactly zero, and the arithmetic and geometric mean coincide. In this sense, approximating the arithmetic mean by the geometric mean is not an additional assumption, but represents another use of the same approximation (namely, constant GDP levels and hence zero growth rates). 
	
	Equation (\ref{secondtolast}) is the unweighted arithmetic mean of $52$ year-over-year log growth rates. Equation (\ref{last}) is the weighted arithmetic mean of $103$ month-over-month log growth rates. Finally, note that the geometric mean is weakly smaller than the arithmetic mean. The two approximations in Equation (\ref{approx2}) hence feature errors of opposite signs: We're underestimating average GDP in year $2$, but also subtract an underestimated version of GDP in year $1$. This setup is beneficial in that it allows the approximation errors to partly offset each other.  
	
	\subsection{State space representation}\label{app:state}
	
	Here we represent the model from Section 3 of the main paper as a linear state space model. Our notation loosely follows \cite{Hamilton1994}. The state equation is given by
	\begin{eqnarray*}
		\xi_w^* &=& F \xi_{w-1}^* + v_w^*,\\
		v_w^* &\stackrel{\text{i.i.d.}}{\sim}& \mathcal{N}(\mathbf{0}_{[103,1]}, Q),
	\end{eqnarray*}
	where $\xi_w^* = \begin{pmatrix} Y_w^*, & Y_{w-1}^*, & \ldots, & Y_{w-102}^* \end{pmatrix}'$ is a $103 \times 1$ vector of weekly log growth rates up until week $w$. $F = \begin{pmatrix} F_1 \\ F_{2} \end{pmatrix}$ is a $103 \times 103$ matrix, where $F_1 = \begin{pmatrix} \phi, & \mathbf{0}_{[1,102]} \end{pmatrix}$ and $F_2 = \begin{pmatrix} I_{102}, & \mathbf{0}_{[102,1]} \end{pmatrix}$, with $I_{102}$ denoting the $102$-dimensional identity matrix, and $\mathbf{0}_{[r, c]}$ denoting an $r \times c$ matrix of zeros. $v_w^*$ is an $103 \times 1$ vector of error terms with mean zero and covariance matrix $Q$, where $Q$ is an $103 \times 103$ matrix with its $[1,1]$ element equal to $\sigma^2_\varepsilon$, and all other elements equal to zero. The observation equation is given by
	\begin{eqnarray*}
		\tilde Y_w^* &=& H'\xi_w^* + \eta_w^*,
	\end{eqnarray*}
	where $$H = \begin{pmatrix} 1,& 0 & \ldots, &  0 \end{pmatrix}$$ is a $103 \times 1$ vector, and $\eta_w^* \stackrel{\text{i.i.d.}}{\sim} \mathcal{N}(0, \sigma^2_\eta)$ is a scalar random variable. Conceptually, $\tilde Y_w^*$ is a noisy version of the weekly  growth rate $Y_w^*$. We can further write $Y_t = \Gamma'\xi_{w_\text{last}(t)}^*,$ where $w_\text{last}(t)$ species the index for the last week of year $t$, and $$\Gamma = \begin{pmatrix} \gamma_1, & \gamma_2,&\ldots, & \gamma_{103} \end{pmatrix}'$$ is a $103 \times 1$ vector collecting the triangle-type aggregation weights defined below Equation (3) in the main paper. With the above setup, standard Kalman filter formulas can be used to compute the optimal $h$-step ahead forecasts of $Y_{t}$, based on the relevant information set $(\tilde Y_w^*)_{w=1}^{w_\text{last}(t)-h}$ consisting of noisy versions of weekly growth rates. In a nutshell, the Kalman filter yields an optimal (`smoothed') estimate of the sequence $(\xi_w^*)_{w=1}^{w_\text{last}(t)-h}$ of state vectors. This estimate is then used to predict the future state vector $\xi_{w_\text{last}(t)}$ and, finally, $Y_t$. See e.g. \cite{Hamilton1994} for an exposition of the relevant computations. 
	
	\subsection{Other parameter choices}\label{app:parameters}
	
\begin{figure}[htbp]
		\centering
		\includegraphics[width=\textwidth]{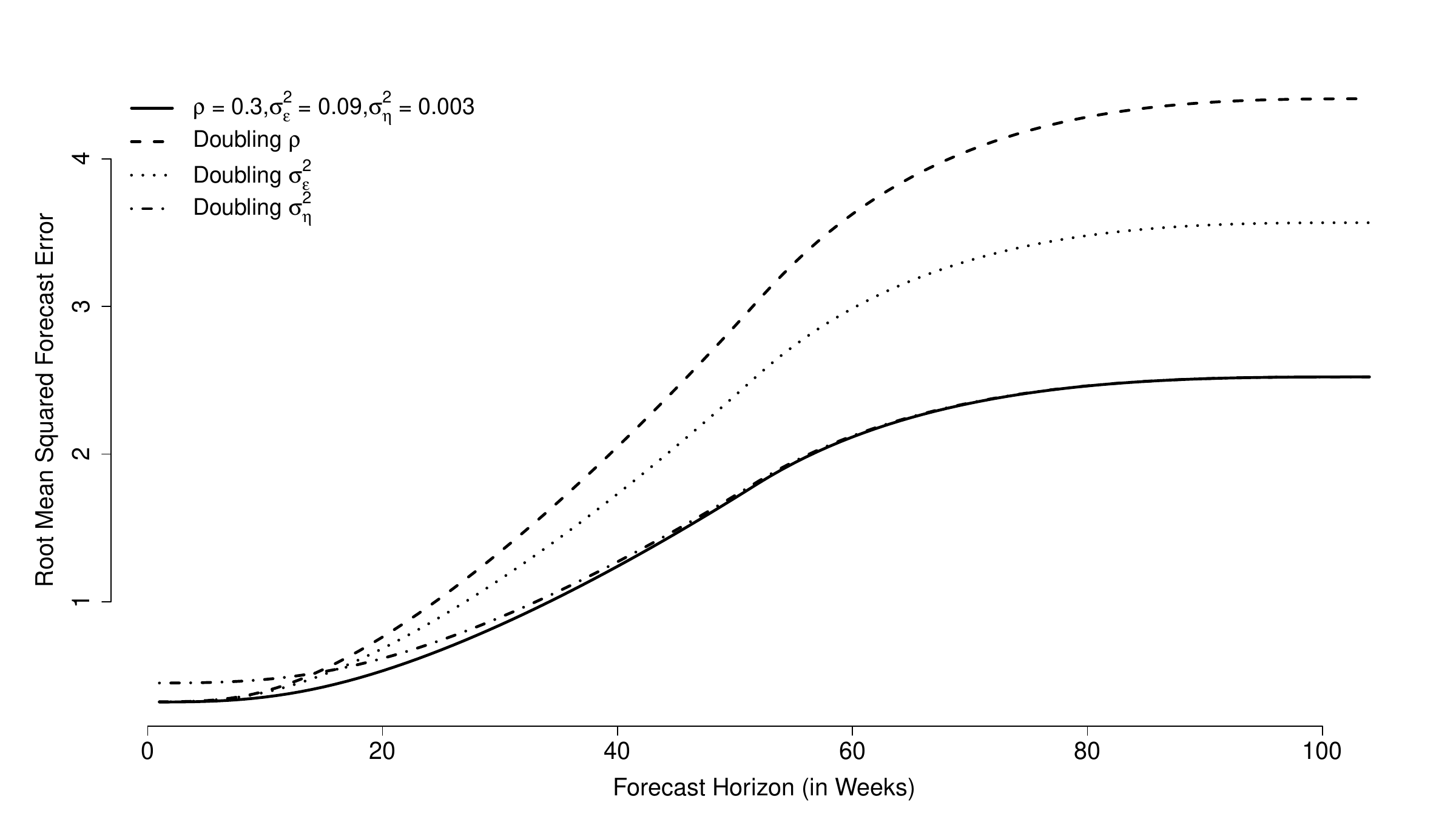}
		\caption{The setup corresponds to Figure 1 from the main paper, and the solid curve is the same as in that figure. The other curves represent different parameter values: In each case, one parameter is doubled (compared to the solid curve), whereas the others are held constant.\label{fig:others}}
	\end{figure}

	Figure \ref{fig:others} explores the model's structural parameters, by changing one parameter at a time. Compared to the default case (solid curve), doubling either $\rho$ (autoregressive persistence) or $\sigma^2_\varepsilon$ (variance of white noise error term in autoregression) increases the RMSFE at all horizons, and tends to yield a steeper increase of the RMSFE across horizons. Doubline $\sigma^2_\eta$ (variance of measurement error) increases the RMSFE at short horizons but has no visible effect at horizons exceeding one year. 
	
	\subsection{Correlation structure of forecast errors}\label{app:cluster}
	
	Consider two forecast errors $e_{t_1, h_1}$ and $e_{t_2, h_2}$, where $e_{t,h}$ is a forecast for year $t$, made $h$ weeks before the end of year $t$. By convention, we let $t_2 \ge t_1$. Furthermore, we assume that $h_1, h_2 \le 104$. Below we provide matrix representations of the two forecast errors. We first introduce the auxiliary matrices $L$ and $A(h)$. The former, which is of dimension $[2,h_1+h_2]$, is given by
	
	$$L = \begin{pmatrix} 
		\mathbf{g}_{[1,h_2]} & \mathbf{0}_{[1,h_1]} \\
		\mathbf{0}_{[1,h_2]} & \mathbf{g}_{[1,h_1]} \\
	\end{pmatrix},$$
	where $\mathbf{g} = \begin{pmatrix} \gamma_1,& \gamma_2,& \ldots,&\gamma_{103},& 0 \end{pmatrix} = \begin{pmatrix} 1, & 2, & \hdots, & 51,& 52, & 51, & \hdots,&2,& 1,&0 \end{pmatrix}/52$ is of dimension $1 \times 104$, and the notation $\mathbf{g}_{[1,r]}$ denotes the first $r$ elements of $\mathbf{g}$. $A(h)$ is an upper triangular $h\times h$ matrix with ones on the main diagonal, and 
	$[i,j]$ element equal to $\rho^{j-i}$ for $j > i$. 
	
	We next introduce a matrix expression for the $2 \times 1$ vector $\begin{pmatrix} e_{t_2,h_2},& e_{t_1,h_1} \end{pmatrix}'$ of forecast errors. This expression allows to assess the errors' correlation structure. In the following expressions, we assume that the matrix $\mathbf{0}_{[r,c]}$ of zeros vanishes whenever $r \le 0$ or $c \le 0$.

	\subsubsection*{Case 1: Both errors refer to the same target year ($t_1 = t_2$)}
	Here the number of relevant weekly shocks (that enter either $e_{t_2,h_2}$ or $e_{t_1,h_1}$ or both) is given by $T^* = \text{max}(h_1, h_2)$. 
	$$\begin{pmatrix} e_{t_2,h_2} \\ e_{t_1,h_1} \end{pmatrix} = L \times  \begin{pmatrix} A(h_2) & \mathbf{0}_{[h_2, T^*-h_2]} \\
		A(h_1) & \mathbf{0}_{[h_1, T^*-h_1]} \\
	\end{pmatrix} \times 
	\begin{pmatrix} \varepsilon_{w_\text{last}(t_2)}^* \\ 
		\varepsilon_{w_\text{last}(t_2)-1}^* \\ \vdots \\
		\varepsilon_{w_\text{last}(t_2)-T^*+1}^*
	\end{pmatrix}$$
	
	\subsubsection*{Case 2: Target years differ by one ($t_2 = t_1 + 1$)}Here $T^*$ (defined as in Case 1) is given by $T^* = \text{max}(h_1+52,h_2)$.
	$$\begin{pmatrix} e_{t_2,h_2} \\ e_{t_1,h_1} \end{pmatrix} = L \times  \begin{pmatrix} A(h_2) & \mathbf{0}_{[h_2, T^*-h_2]} &  \\
		\mathbf{0}_{[h_1, 52]} & A(h_1) &\mathbf{0}_{[h_1, T^*-52-h_1]} \\
	\end{pmatrix} \times 
	\begin{pmatrix} \varepsilon_{w_\text{last}(t_2)}^* \\ 
		\varepsilon_{w_\text{last}(t_2)-1}^* \\ \vdots \\
		\varepsilon_{w_\text{last}(t_2)-T^*+1}^*
	\end{pmatrix}$$
	
	\subsubsection*{Case 3: Target years differ by two or more ($t_2 \ge t_1 + 2$)} Recalling that $h_1,h_2 \le 104$, we have $T^* = (t_2-t_1)\times 52 + h_1$, and		
	$$\begin{pmatrix} e_{t_2,h_2} \\ e_{t_1,h_1} \end{pmatrix} = L \times  \underbrace{\begin{pmatrix} A(h_2) & \mathbf{0}_{[h_2,T^*-h_2]} &  \\
			\mathbf{0}_{[h_1,T^*-h_1]} & A(h_1) \\
	\end{pmatrix}}_{\underline{A}} \times 
	\begin{pmatrix} \varepsilon_{w_\text{last}(t_2)}^* \\ 
		\varepsilon_{w_\text{last}(t_2)-1}^* \\ \vdots \\
		\varepsilon_{w_\text{last}(t_2)-T^* +1}^*
	\end{pmatrix}$$		
	
	We next illustrate the computation of the errors' covariance matrix for Case 3. Recalling the assumption of Gaussian IID errors $\varepsilon^*_m$, with $\text{Var}(\varepsilon^*_m) = \sigma^*_\varepsilon,$ we get
	$$\text{Var}\begin{pmatrix} e_{t_2,h_2} \\ e_{t_1,h_1} \end{pmatrix} = \sigma^2_\varepsilon~L~\underline{A}~\underline{A}'L'.$$ By examining the definitions of $L$ and $\underline{A}$, it becomes clear that $\text{Var}\begin{pmatrix} e_{t_2,h_2} \\ e_{t_1,h_1} \end{pmatrix}$ is a diagonal matrix in Case 3. If $h_2 \le 52$, the same finding obtains in Case 2. Otherwise (in Case 2 with $h_2 \ge 53$, or in Case 1), we generally obtain a non-diagonal covariance matrix for $\begin{pmatrix} e_{t_2,h_2} \\ e_{t_1,h_1} \end{pmatrix}$.

	\section{Additional empirical results}
	
	\label{sec:empos}
	
	\subsection{Forecast calibration}
	
\begin{figure}[htp]
		\centering
		\begin{tabular}{ccc}
			\textit{German GDP} & \textit{US GDP} & \textit{US Inflation} \\ \toprule
			&&\\
			\multicolumn{3}{c}{Gaussian}\\[-.3cm]
			\includegraphics[width=.25\textwidth]{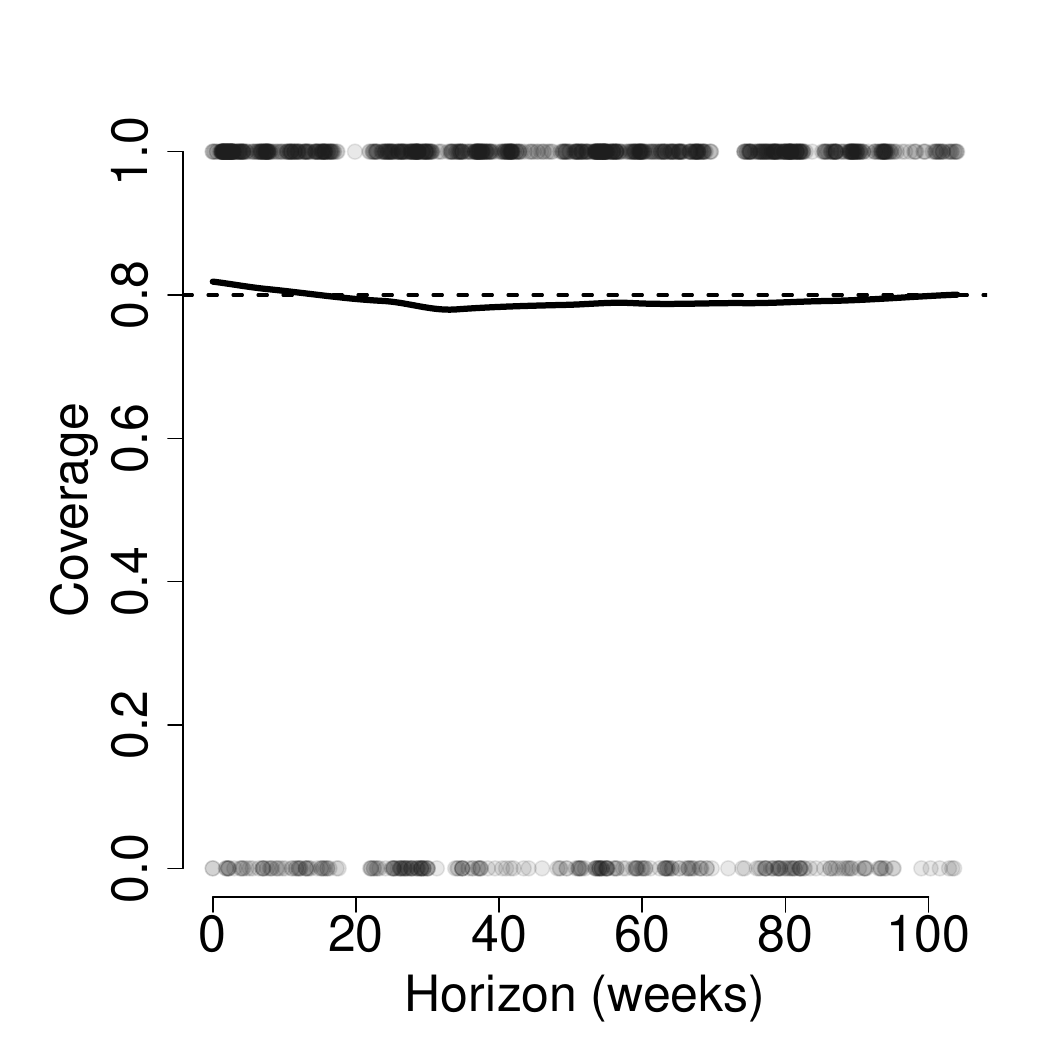} & 
			\includegraphics[width=.25\textwidth]{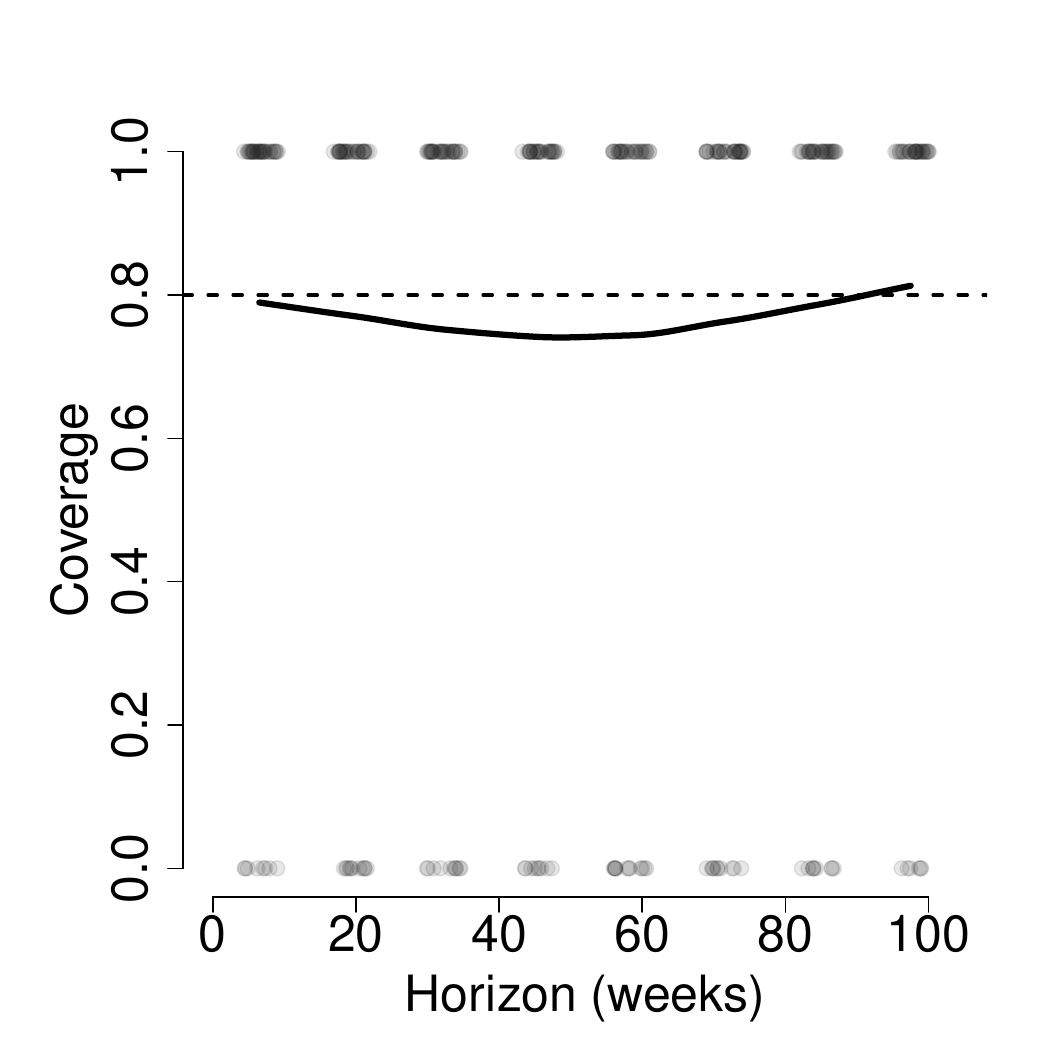} & 
			\includegraphics[width=.25\textwidth]{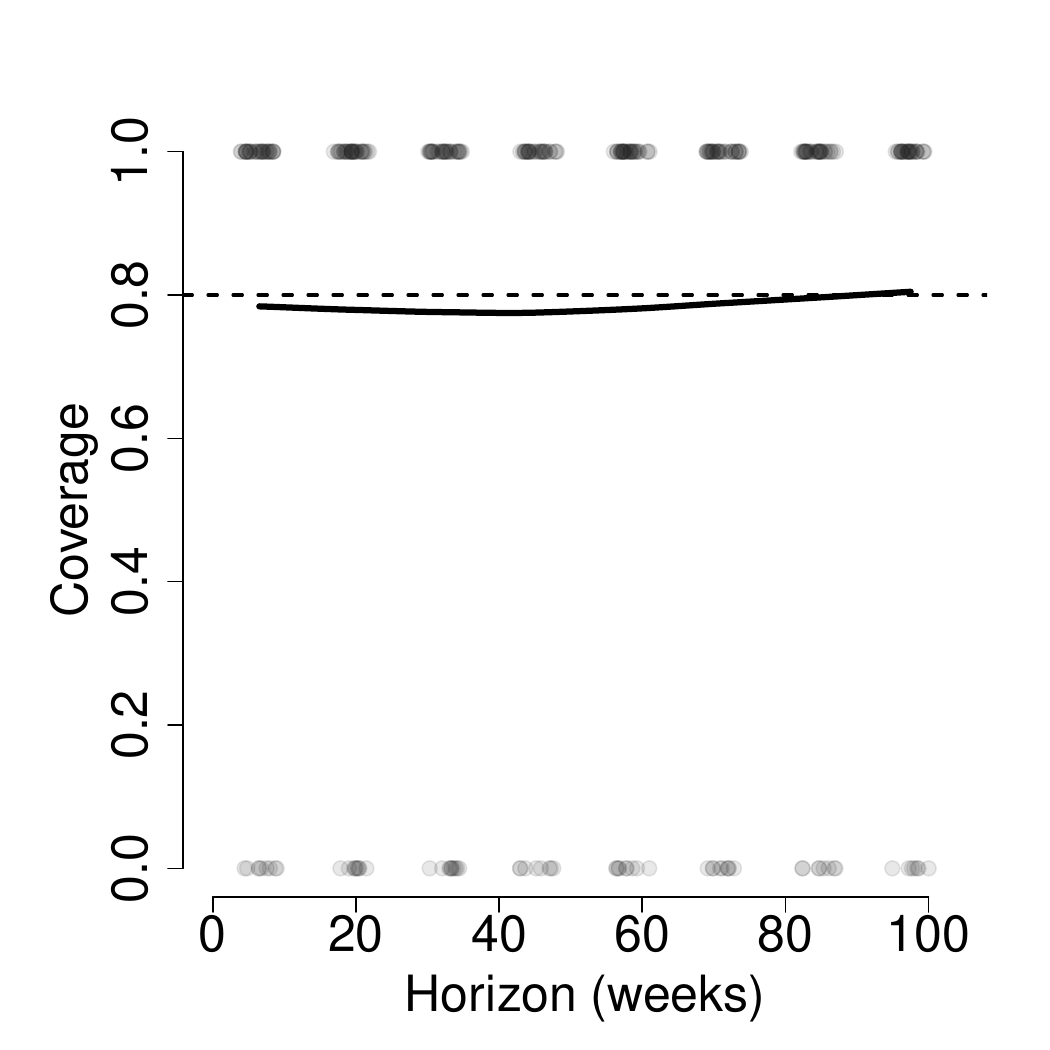} \\[.2cm]	
			\multicolumn{3}{c}{Decomposition}\\[-.3cm]
			\includegraphics[width=.25\textwidth]{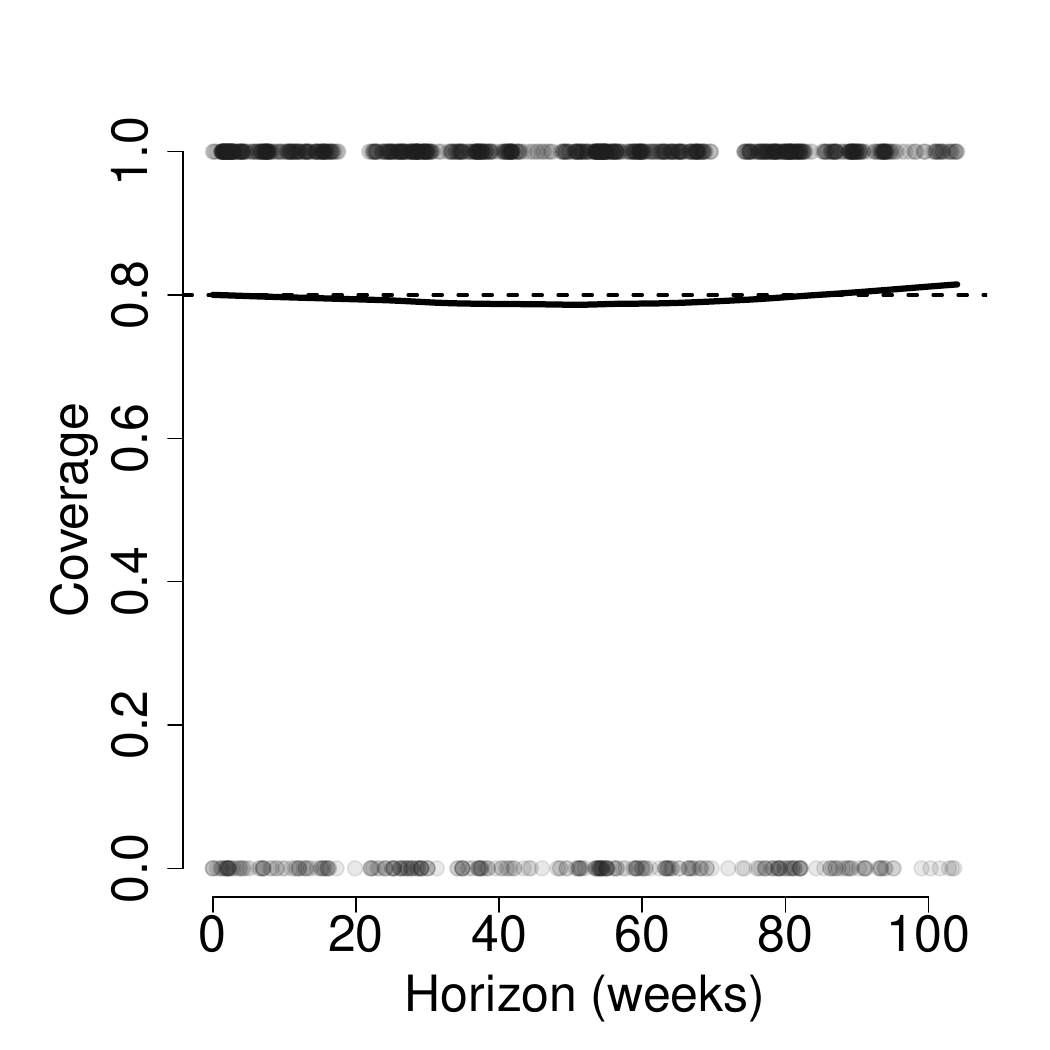} & 
			\includegraphics[width=.25\textwidth]{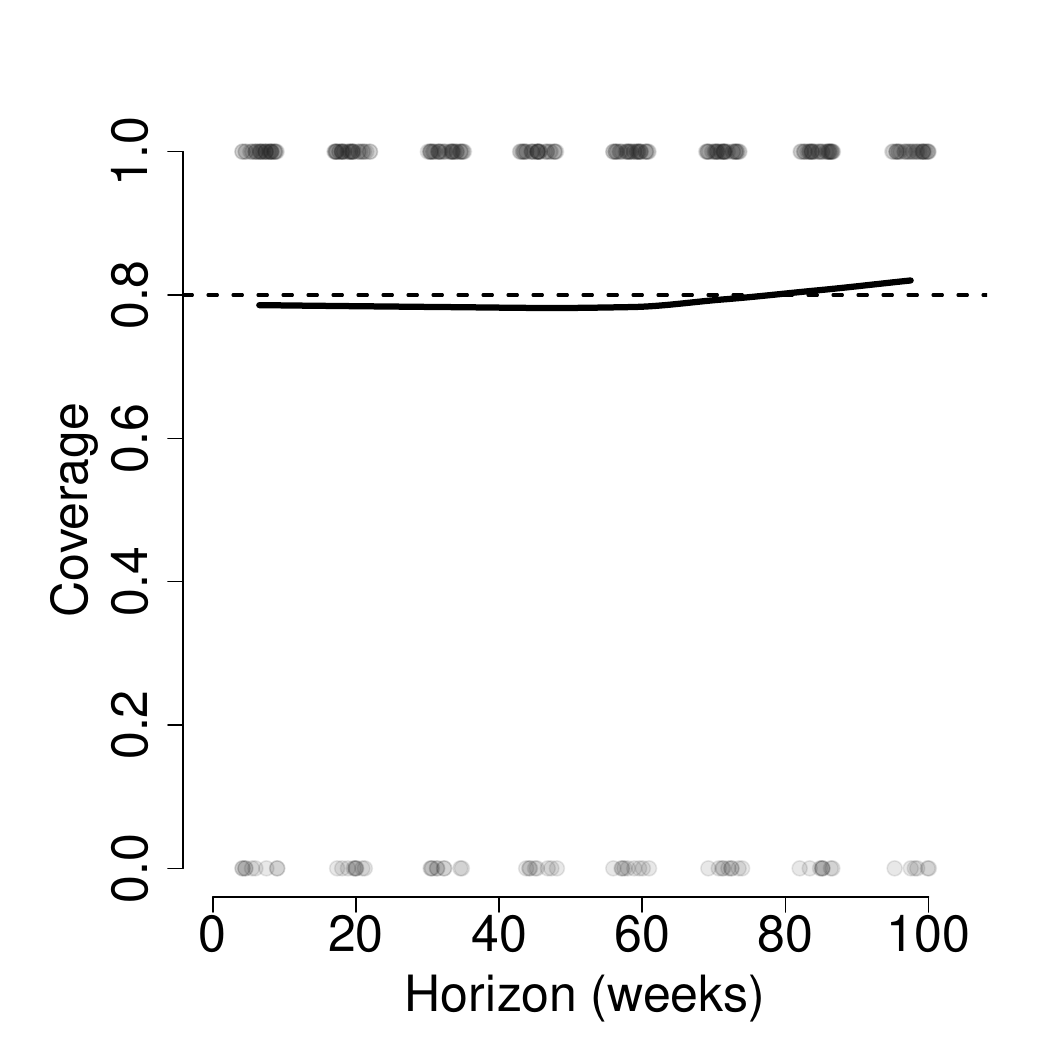} & 
			\includegraphics[width=.25\textwidth]{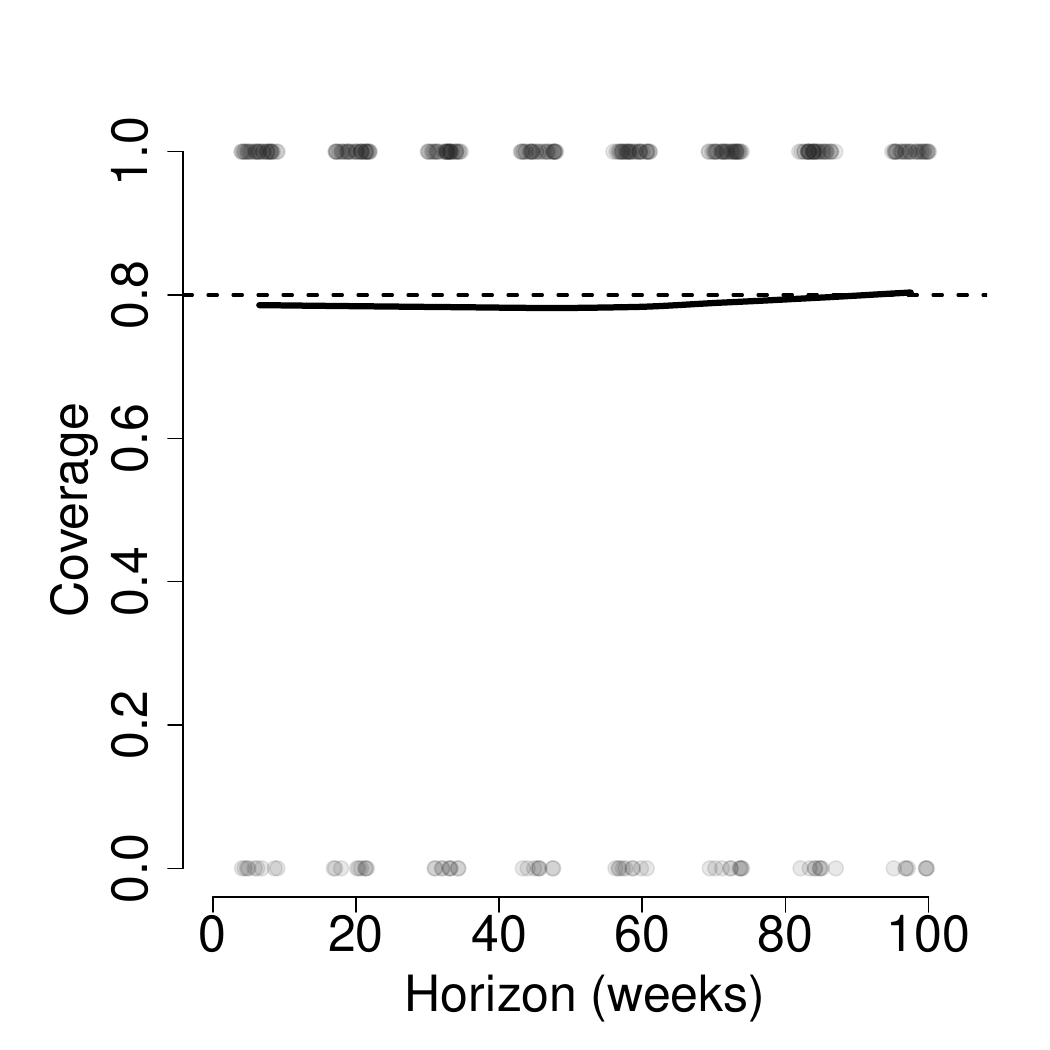} \\[.2cm]	
			\multicolumn{3}{c}{Flexible}\\[-.3cm]
			\includegraphics[width=.25\textwidth]{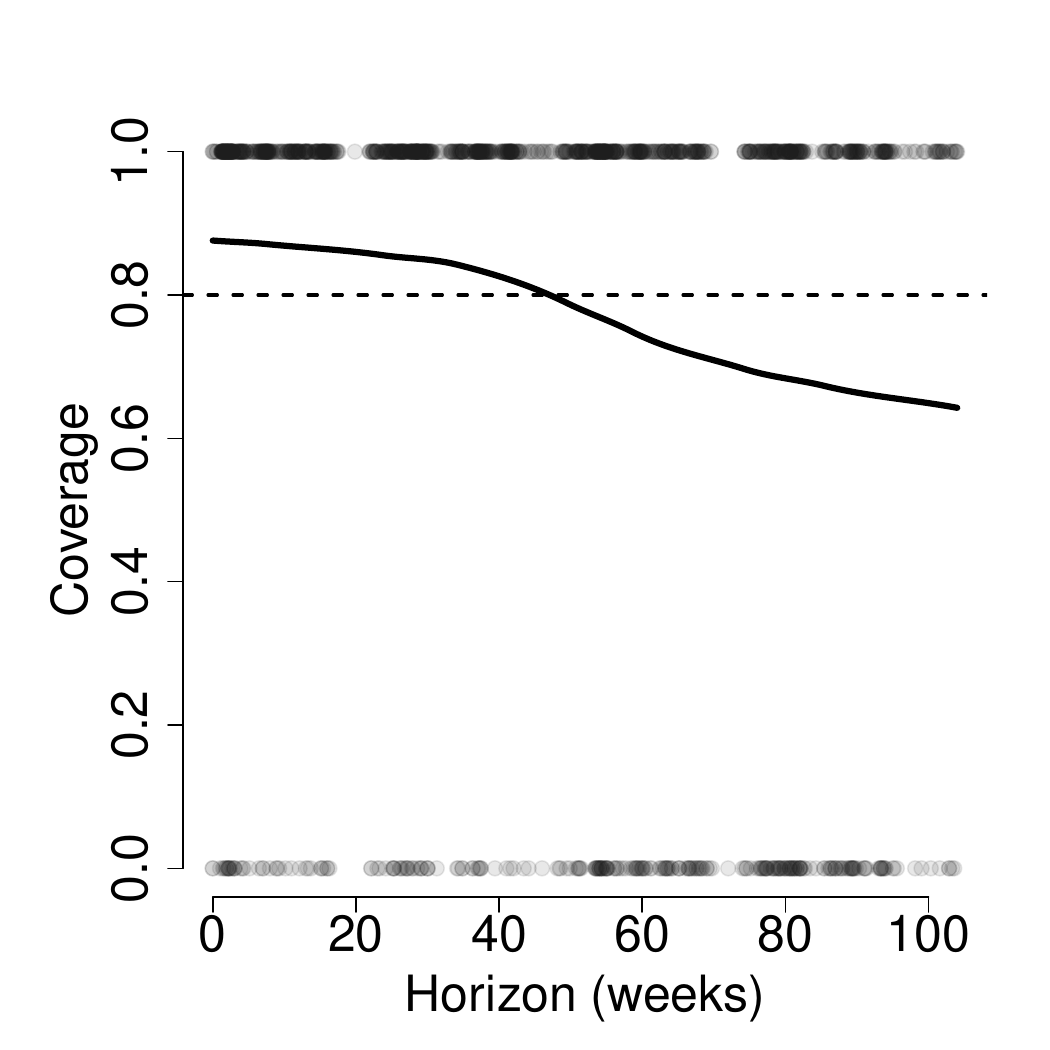} & 
			\includegraphics[width=.25\textwidth]{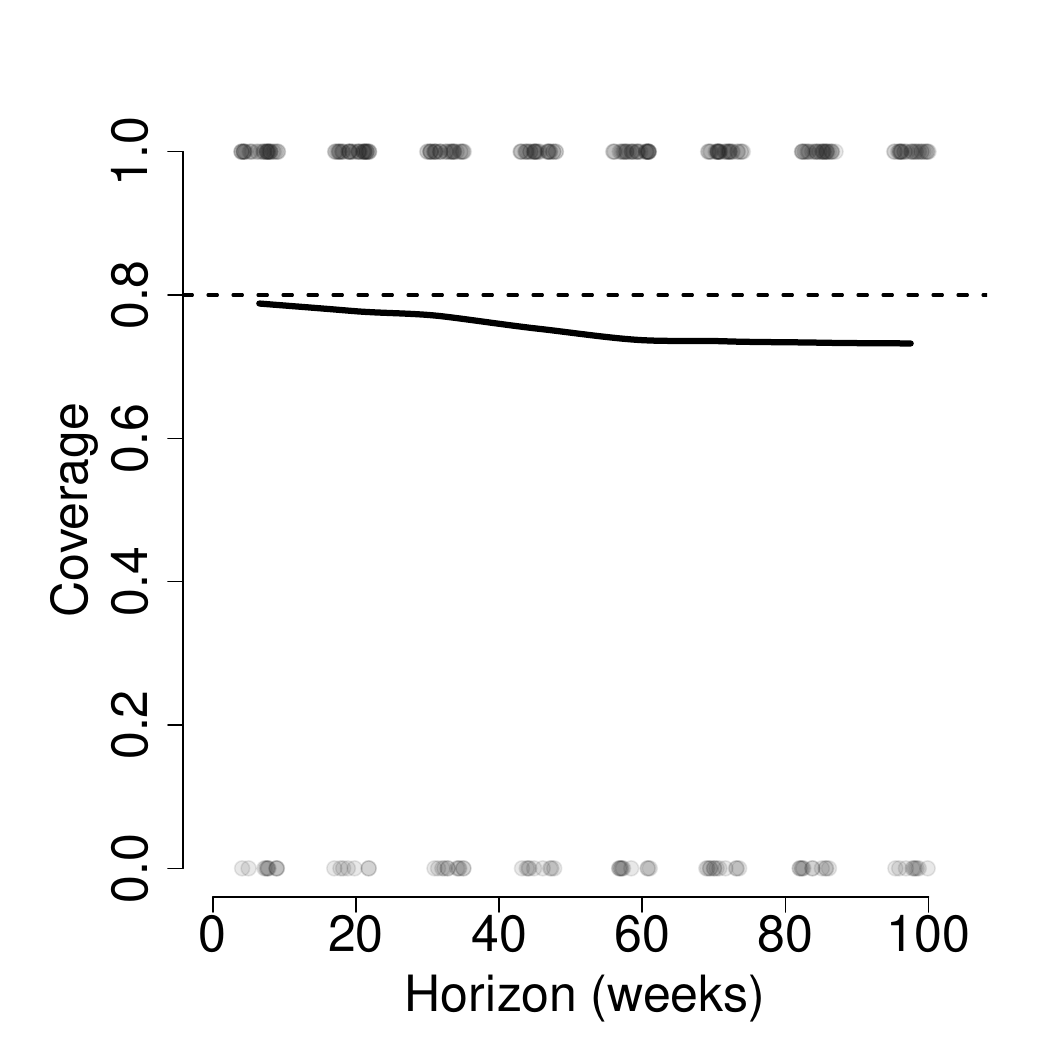} & 
			\includegraphics[width=.25\textwidth]{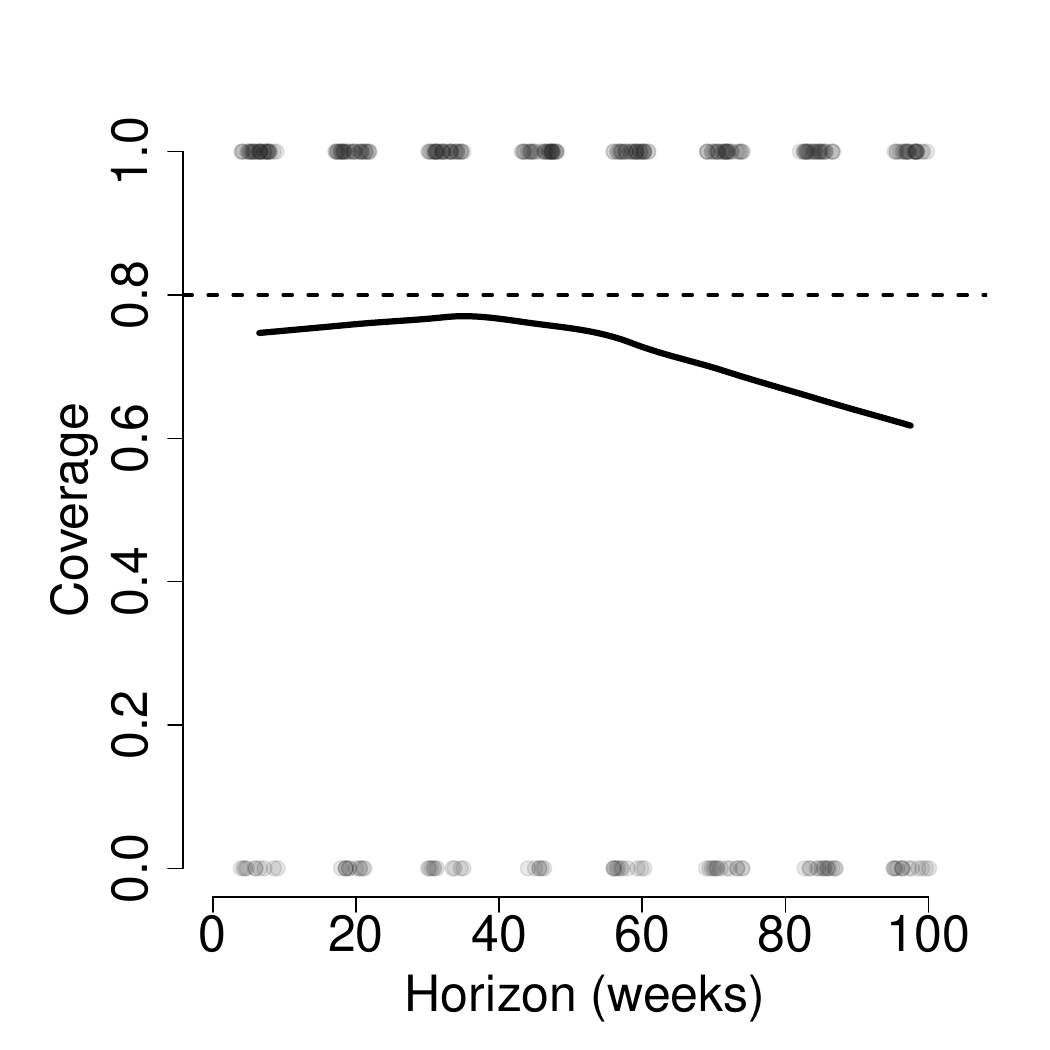} \\[.2cm]	
			\multicolumn{3}{c}{Combination}\\[-.3cm]
			\includegraphics[width=.25\textwidth]{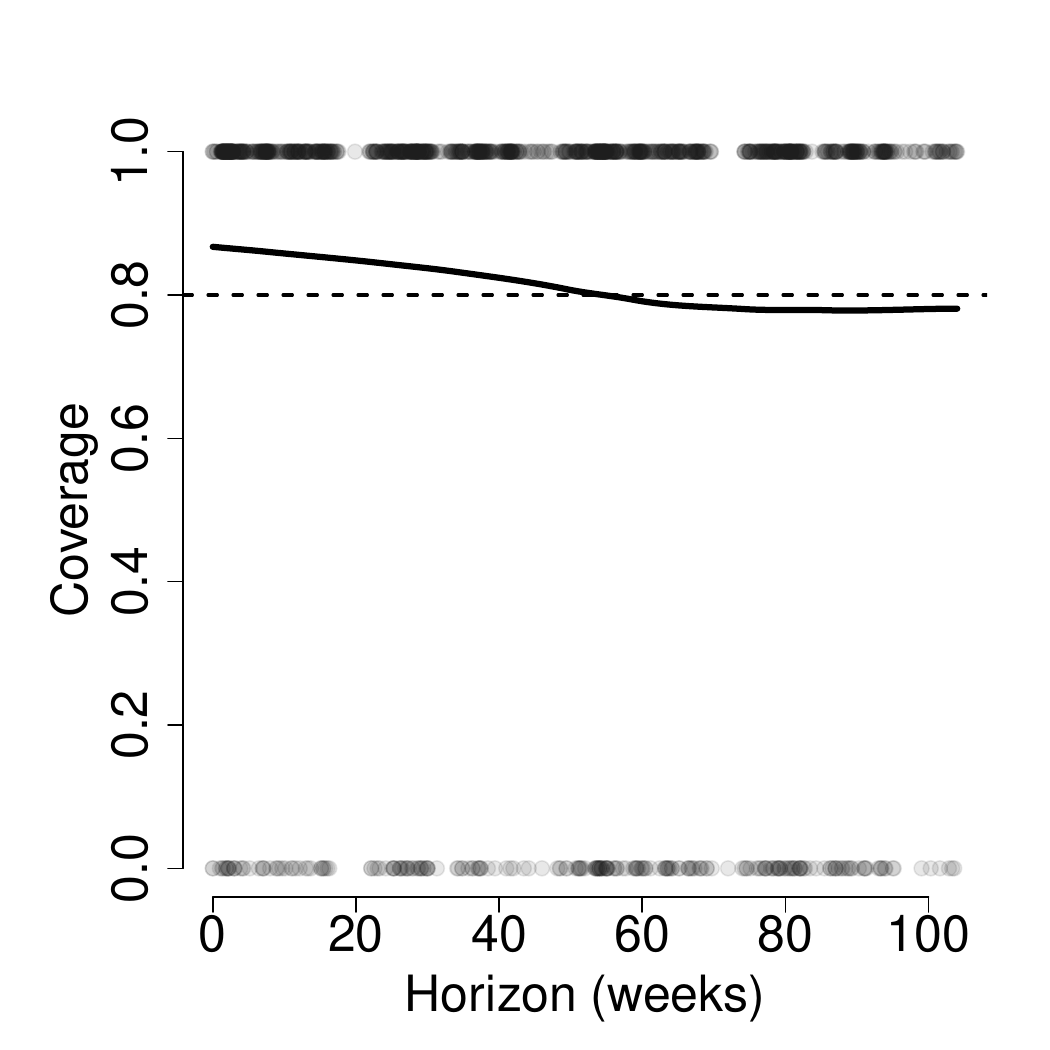} & 
			\includegraphics[width=.25\textwidth]{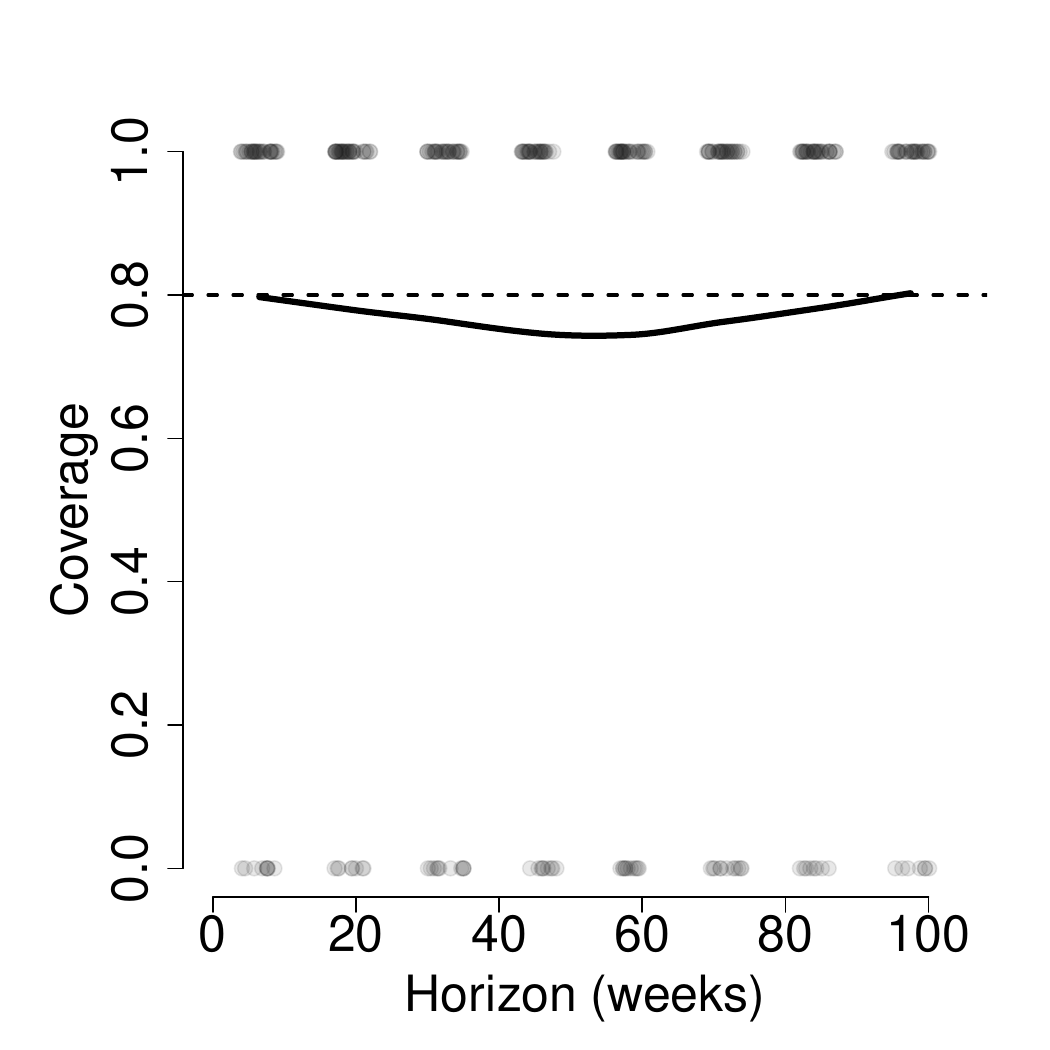} & 
			\includegraphics[width=.25\textwidth]{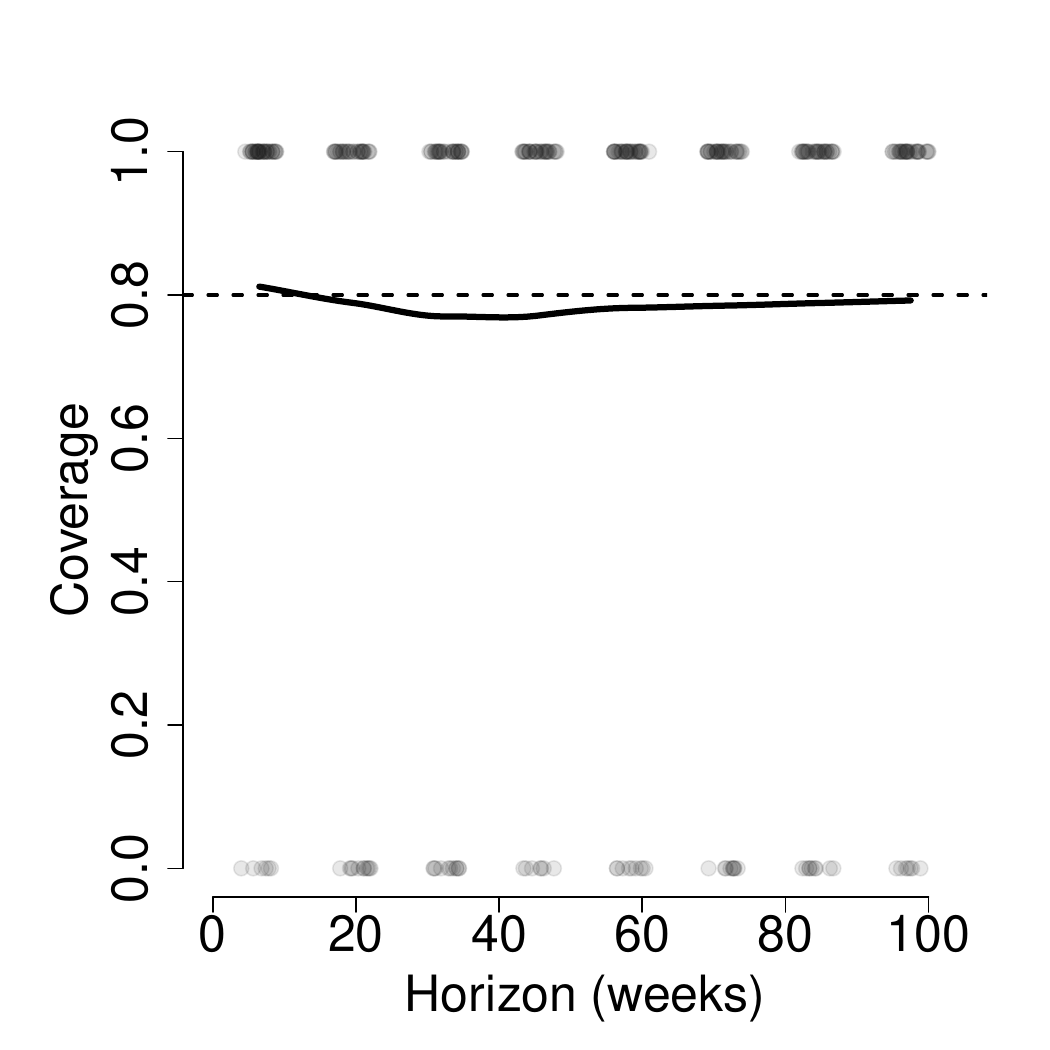} \\ \bottomrule
		\end{tabular} 	
		\caption{Coverage of the prediction interval (vertical axis; 0 = no, 1 = yes) at various horizons (horizontal axis; jittering used for better display). Dots represent individual forecast/outcome pairs. Solid line shows nonparametric estimate of coverage probability as a function of the horizon. Dashed horizontal line indicates the target coverage probability of $80\%$.\label{fig:covh}}
	\end{figure}

	Figure \ref{fig:covh} presents results on the coverage of the prediction intervals, evaluated separately across forecast horizons. 
	
	\subsection{Diebold-Mariano type testing}
	\label{sec:dm}
	
	In Section \ref{sec:dm}, we provide additional discussion and results on Diebold-Mariano type testing for equal predictive ability of postprocessing methods versus the SPF `histograms'. 
	
	\subsubsection{Implementation variants}
	
	We first repeat the Diebold-Mariano (DM) tests reported in Figure 3 of the main paper, using various test implementations. The DM test refers to the null hypothesis that $\mathbb{E}(d_{t,h}) = 0,$ i.e., that the expected difference between the interval scores of two methods is zero.\footnote{In more detail, let $d_{t,h}$ denote the difference in interval scores between the two methods of interest: Combination of postprocessing methods, henceforth abbreviated by `comb.', and the average of the histogram type forecasts from the Survey of Professional Forecasters, henceforth `SPF'. For a given target year $t$ and forecast horizon $h$, this difference is given by 
		$$d_{t,h} = \text{IS}(l_{t,h}^\text{comb.}, u_{t,h}^\text{comb.}, y_t)- \text{IS}(l_{t,h}^\text{SPF}, u_{t,h}^\text{SPF}, y_t),$$
		where $l_{t,h}^j$ denotes the lower end of the prediction interval produced by method $j \in \{\text{comb.}, \text{SPF}\}$, for target year $t$ and horizon $h$. The definition of the upper end $u_{t,h}^j$ of the interval is analogous, and $y_t$ denotes the realizing observation. The interval score is defined in Equation (6) of the main paper.} Here $t$ denotes the target year, and $h$ denotes the forecast horizon (measured in weeks). The DM tests are conducted separately across horizons $h \in \{6.5, 19.5, \ldots, 97.5\}$, where the choice of horizons conforms to the timing of the SPF.
	We thus seek to test the null hypothesis based on a sample $(d_{t,h})_{t=1}^{n_h}$, using a $t$-statistic of the form $$\frac{\bar{d}_h}{\widehat V(\bar d_h)},$$ where $\bar d_h = n_h^{-1} \sum_{t=1}^{n_h} d_{t,h}$ is the sample average of the loss differences, and $\hat V$ denotes an estimate of the variance of the sample average. Due to data availability, the size $n_h$ of the evaluation sample ranges between $36$ and $42$, depending on the horizon $h$. Common choices of $\hat V$ depend on estimates or assumptions relating to the time series dependence (specifically, autocorrelation) in $(d_{t,h})_{t=1}^{n_h}$. To discuss this aspect, note that the loss differences $d_{t,h}$ are observed annually, i.e., the target years for the observations $d_{t,h}$ and $d_{t+1, h}$ are one year apart. Since four of the eight forecast horizons $h$ we consider are less than one year, the informal benchmark setup suggested by \citet[Section 1.1]{DieboldMariano1995} corresponds to a truncation lag of zero (i.e., no consideration of autocorrelation) for these horizons, and a truncation lag of one (i.e., consideration of autocorrelation up to lag one) for the remaining four horizons.\footnote{ \cite{DieboldMariano1995} denote the difference in losses by $d_t$, and the forecast horizon by $k$. Unlike in our notation, $t$ refers to an arbitrary time scale (e.g., monthly, quarterly, or annually), and the forecast horizon $k$ is measured on the same scale as $t$. Based on properties of optimal $k$-step ahead forecast errors, \citeauthor{DieboldMariano1995} propose using $(k-1)$ as a benchmark choice of truncation lag.} Alternatively, the truncation lag can be determined according to data-based procedures that are popular in the literature on covariance matrix estimation in the presence of time dependence.\\
	
	We consider the following DM test implementations:
	\begin{enumerate}
		\item \textbf{sandwich}: The estimator implemented in the function \textsf{NeweyWest} of the R package \textsf{sandwich} \citep{Zeileis2004,ZeileisEtAl2020}, implementing the data-based choice of truncation lag described in \cite{NeweyWest1994}. We compare the resulting $t$-statistic to standard normal critical values. This choice is the same as in Figure 3 of the main paper. 
		\item \textbf{CM13}: An implementation variant that performs well in Monte Carlo simulations by \cite{ClarkMcCracken2013} on DM testing under squared error loss. The test uses standard normal critical values. Variance estimation is based on a rectangular kernel, with the truncation lag chosen according to the \citeauthor{DieboldMariano1995} benchmark setup mentioned above. Furthermore,  the finite sample adjustment proposed by \cite{HarveyEtAl1997} is applied to the test statistic. (Referring to the authors of the latter study, \citeauthor{ClarkMcCracken2013} denote this implementation variant by `HLN'.)
		\item \textbf{EWC}: A test statistic using the equally weighted cosine (EWC) variance estimator proposed by \cite{LazarusEtAl2018}. We compare the test statistic against critical values from a $t$ distribution with $\nu$ degrees of freedom, where $\nu$ is an implementation parameter that is set according to the rule in \citet[Equation 4 and Footnote 3]{LazarusEtAl2018}. For our choices of sample size $n_h$, the rule prescribes to set $\nu = 4$. 
		\item \textbf{IID}: A standard $t$-test that does not account for autocorrelation. We compare the test statistic against critical values from a $t$ distribution with $n_h-1$ degrees of freedom. This test serves as a simple baseline, and illustrates the impact of ignoring autocorrelation. 
	\end{enumerate}
	
	While this list of DM implementation variants is quite diverse, it is far from exhaustive. See \cite{CoroneoIacone2020} and \cite{CoroneoEtAl2024} for further options, including the use of bootstrap techniques and different asymptotic frameworks.\\
	
\begin{figure}
		\centering
		\begin{tabular}{cc}
			\multicolumn{2}{c}{\textbf{Diebold-Mariano $p$-values}}\\[.2cm]
			\textit{US GDP} & \textit{US Inflation} \\ 
			& \\ [-1cm]
			\includegraphics[width = .4\textwidth]{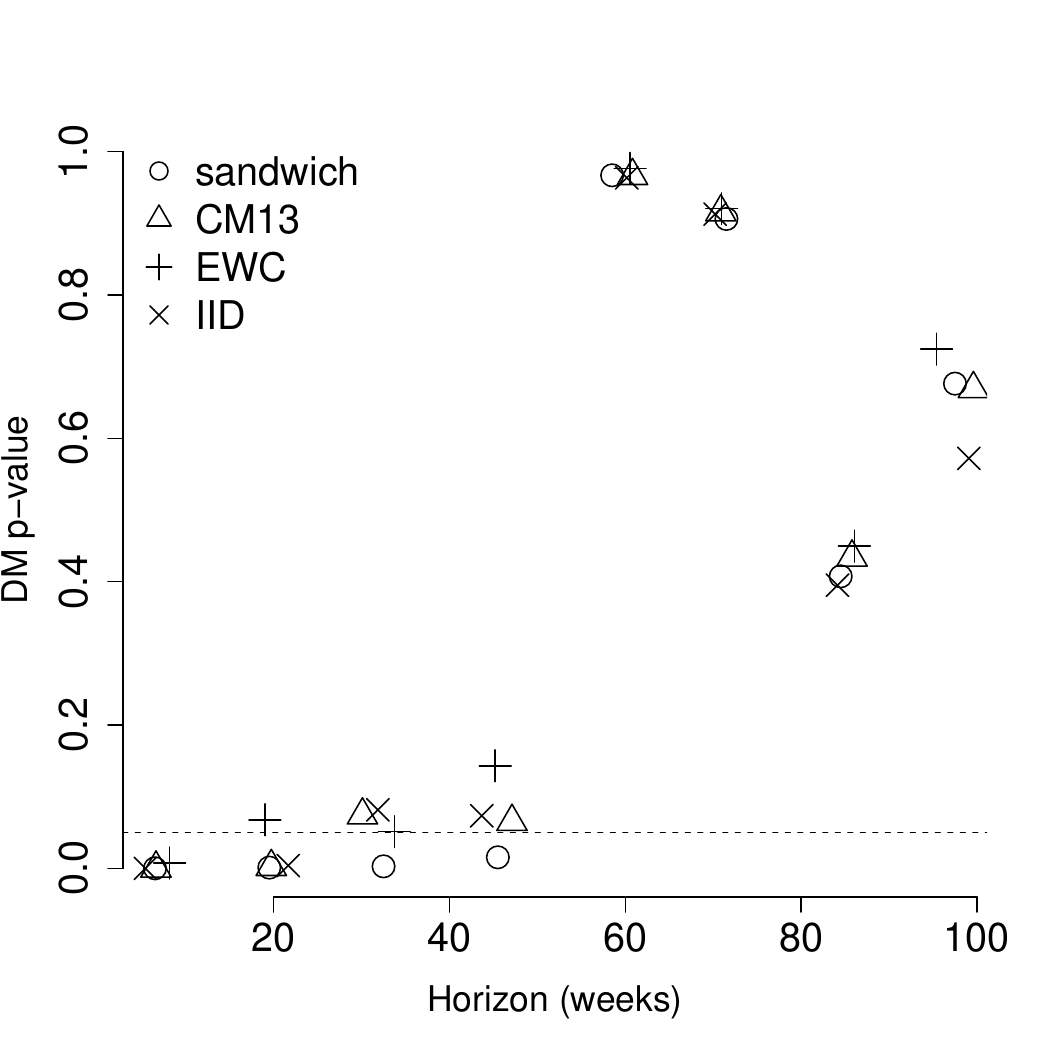} & 
			\includegraphics[width = .4\textwidth]{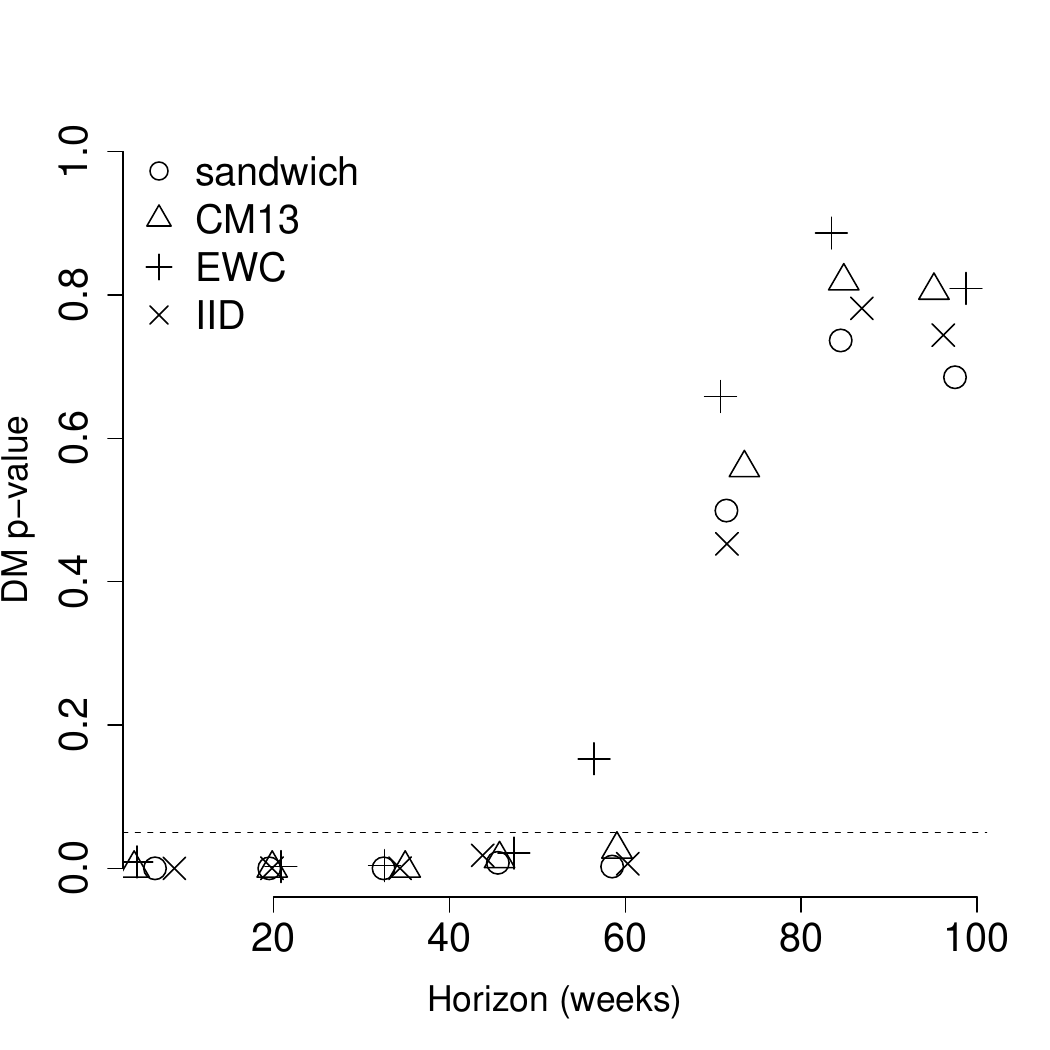} \\
			&\\
			\multicolumn{2}{c}{\textbf{First-order autocorrelation of loss differences}}\\[.2cm]
			\textit{US GDP} & \textit{US Inflation} \\ 
			& \\ [-1cm]
			\includegraphics[width = .4\textwidth]{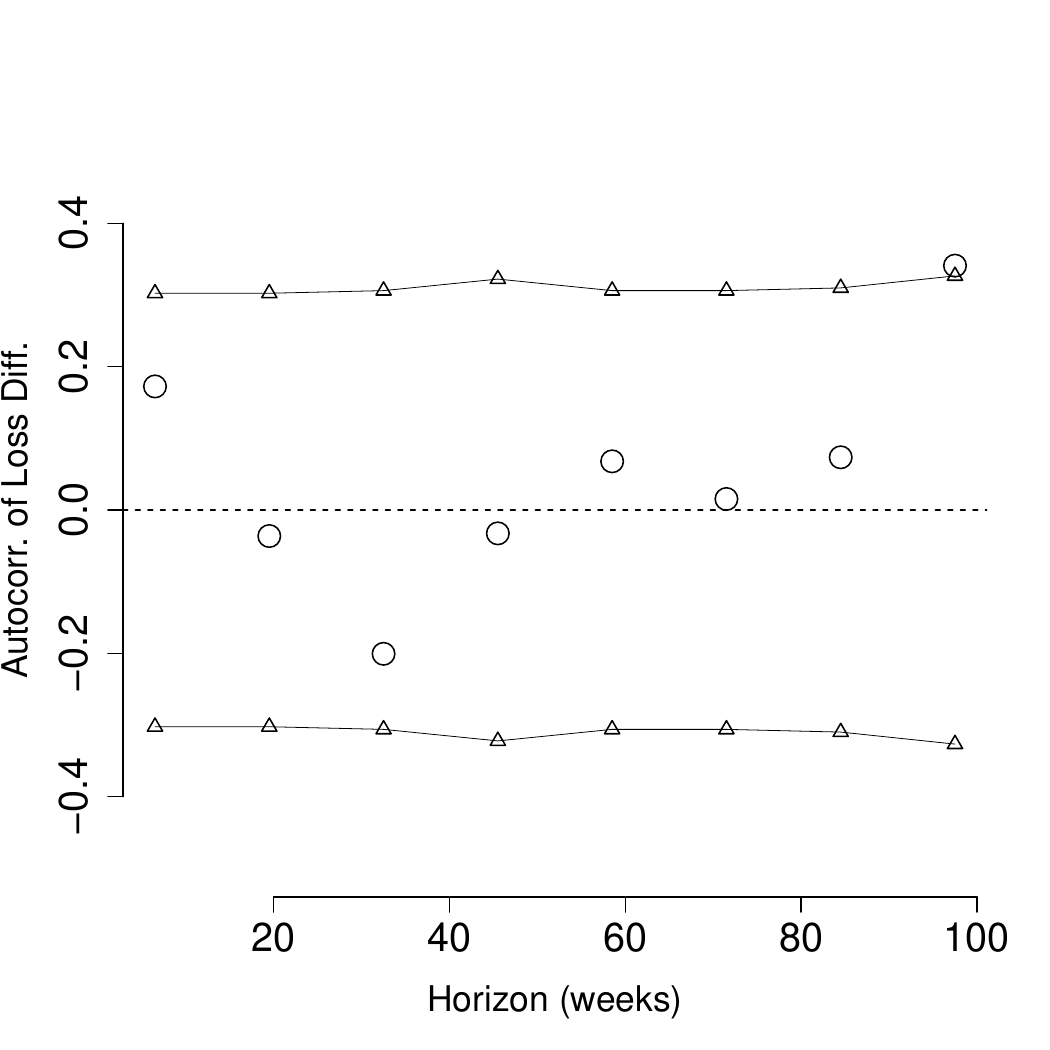} & 
			\includegraphics[width = .4\textwidth]{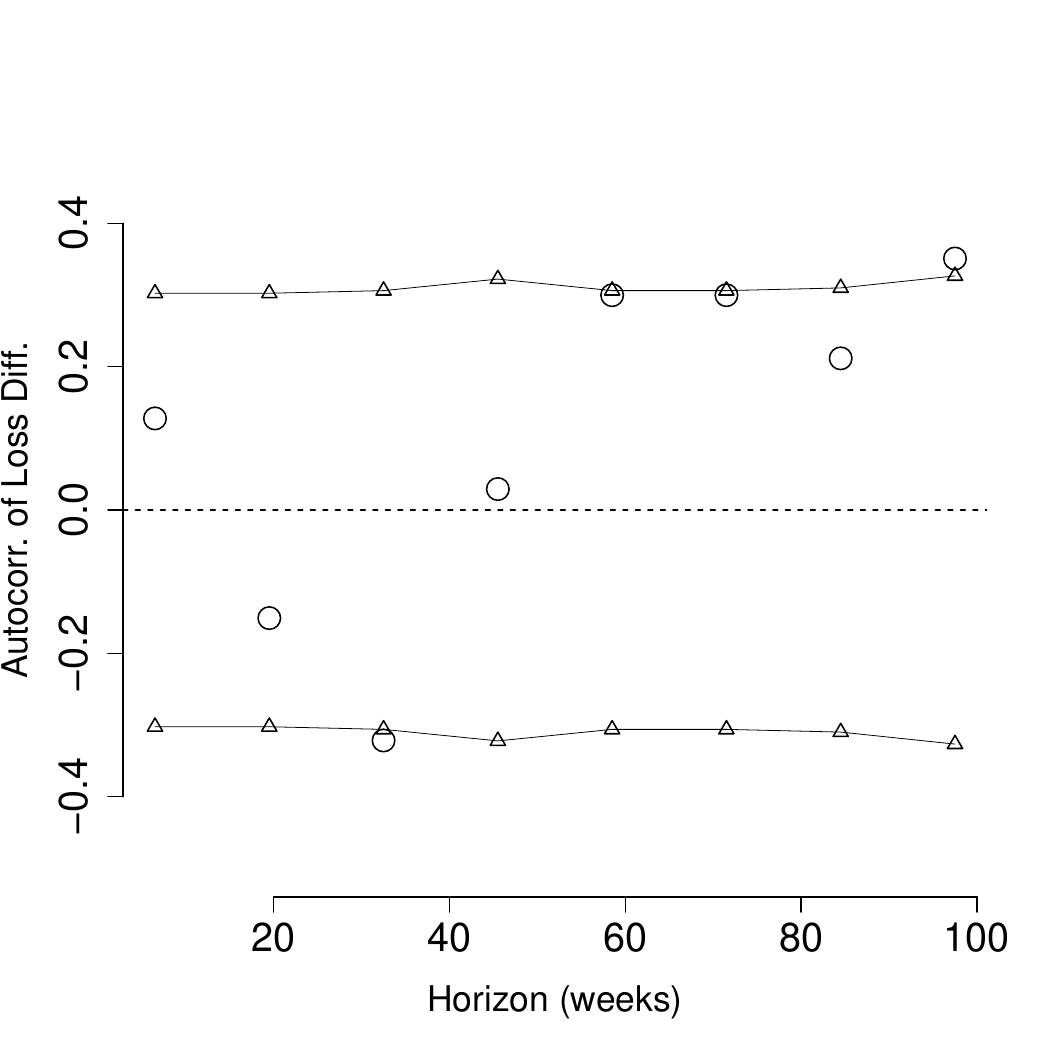} \\ 
		\end{tabular}
		\caption{Top row: Diebold-Mariano $p$-values obtained from different test implementations (two-sided tests). Horizontal location of points is jittered for better display. Horizontal line marks a $p$-value of five percent. Bottom row: Circles mark first-order autocorrelation of loss differences $d_{t,h}$. Triangles (connected by lines) mark asymptotic critical values 
			for the null hypothesis of zero autocorrelation \citep[see][Section 4.4.1]{Luetkepohl2005} at the $5\%$ level. \label{tab:dmpvals}}
	\end{figure}

	The top row of Figure \ref{tab:dmpvals} shows the different $p$-values for GDP (left panel) and inflation (right panel). The four test implementations generally show a high level of agreement. The main differences occur for GDP at horizons $h \in \{19.5, 32.5, 45.5\}$. Even for these settings, however, all tests reject the null at the ten percent level except in one case (EWC, $h = 45.5$). Generally, there is a tendency for EWC to be more conservative (i.e., produce larger $p$-values) than the other tests.\\
	
	As noted, the four test implementations differ in their handling of potential autocorrelation in the loss differences. The bottom row of Figure \ref{tab:dmpvals} presents empirical evidence on autocorrelation, by showing the empirical first-order autocorrelation coefficients of the series $(d_{t,h})_{t=1}^{n_h}$, for different horizons $h$. For both variables, these coefficients attain small positive or even negative values at the four shortest horizons. They are somewhat larger at the four longer horizons, but remain moderate (below $0.36$) even there. The figures also show that there is substantial sampling uncertainty, with few of the autocorrelation coefficients being significantly different from zero at the five percent level. As a consequence, it is not obvious which DM test variant is most appropriate in terms of size and power in the current situation. 
	
	\subsubsection{Performance comparisons on average across horizons}
	
\begin{figure}[!h]
		\centering
		\begin{tabular}{cc}
			\textit{US GDP} & \textit{US Inflation} \\ 
			& \\[-1cm]
			\includegraphics[width = .4\textwidth]{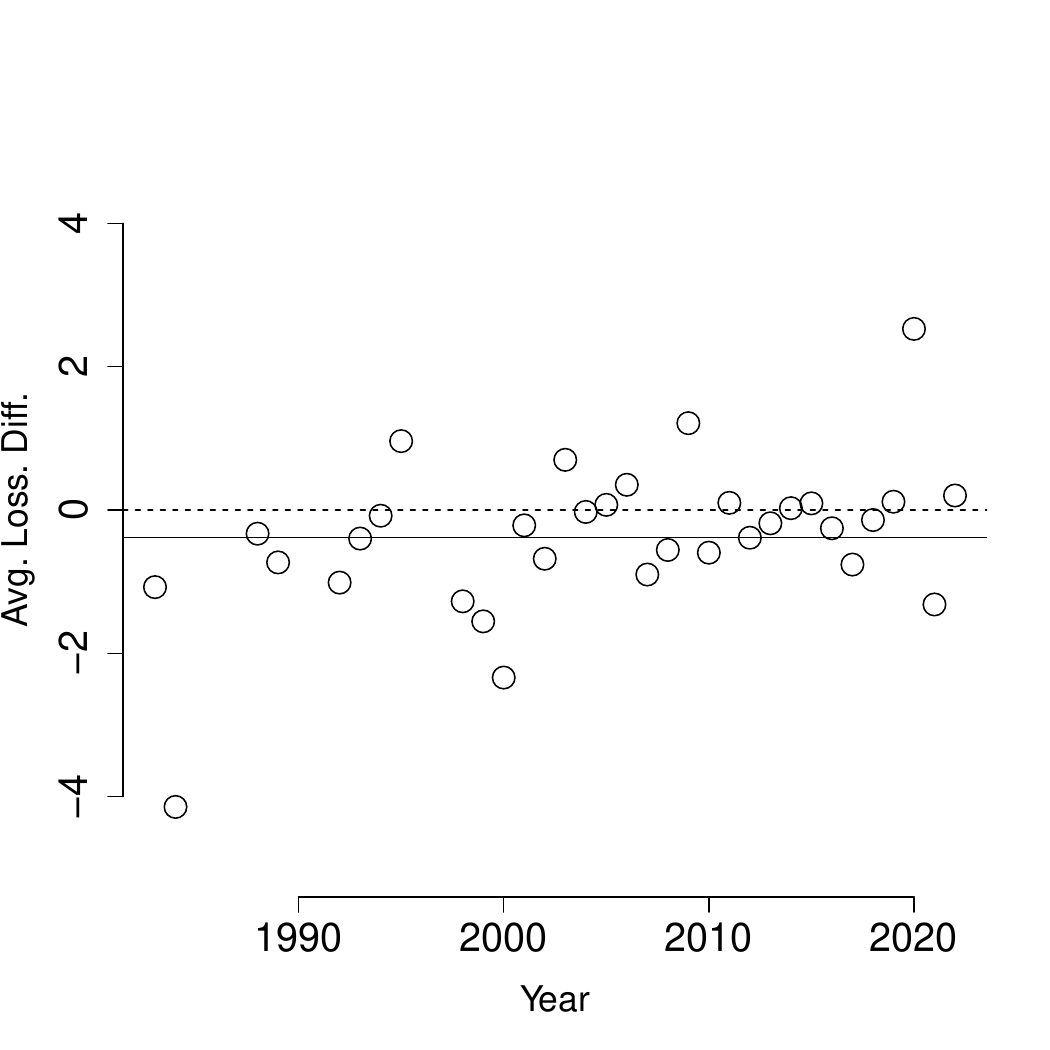} & 
			\includegraphics[width = .4\textwidth]{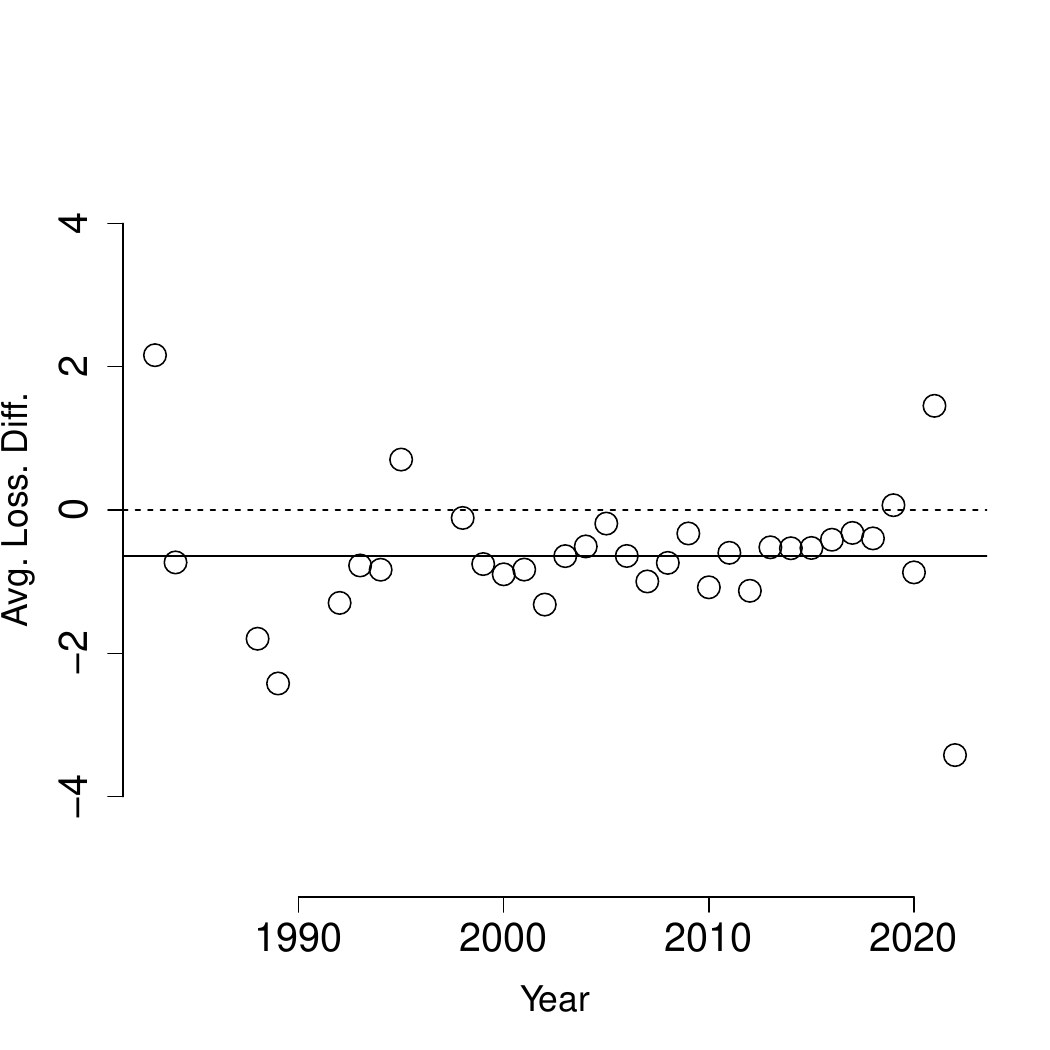} \\ 
		\end{tabular}
		\caption{Average loss differences across horizons, for various target years $t$. Dashed horizontal line marks zero. Solid horizontal line marks the average loss difference over time. \label{avg_ld}}
	\end{figure}

	Instead of conducting DM tests separately for each horizon $h$, it is also possible to evaluate forecasting performance jointly across all horizons. Similar to \cite{Quaedvlieg2021}, one strategy for doing so is to consider average forecast performance across horizons. To this end, let $\bar d_t = \frac{1}{8}\sum_{h \in \mathcal{H}} d_{t,h},$ where the set $\mathcal{H} = \{6.5, 19.5, \ldots, 97.5\}$ comprises the eight forecast horizons of interest. Figure \ref{avg_ld} shows the time series of $d_t$ for GDP (left panel) and inflation (right panel). The loss differences are defined such that positive values are in favor of the SPF, whereas negative values are in favor of the combination.\\
	
	We can now test the null hypothesis that $\mathbb{E}(\bar d_{t}) = 0$ using a DM test like the ones considered above. (Alternatively, \cite{Quaedvlieg2021} considers bootstrap techniques for testing this hypothesis.) In order to implement the test, we require complete data on loss differences at all eight horizons $h$, for any given year $t$. This requires us to remove years for which one or more horizons are missing, so that our final data set includes $33$ observations (i.e., years).\\
	
	Table \ref{tab:dm_av} presents the results of this analysis. For GDP, the EWC test variant yields a $p$-value of $19.2\%$, whereas the other variants attain $p$-values around five percent. For inflation, all four test variants yield very small $p$-values below $0.5\%$. In all cases, rejections are due to negative $t$-statistics (i.e., in favor of the combination method).

	\begin{table}
		\centering
		\begin{tabular}{lcc}
			& \textit{US GDP} & \textit{US Inflation} \\ \midrule
			\textbf{sandwich} & $0.050$  & $0.000$\\
			\textbf{CM13} & $0.047$ & $0.000$\\
			\textbf{EWC} & $0.192$ & $0.000$\\
			\textbf{IID} & $0.055$ & $0.001$\\ \midrule
		\end{tabular}
		\caption{$p$-values for the null hypothesis that $\mathbb{E}(\bar d_t) = 0$. Two-sided tests. Rows represent four DM test implementations. \label{tab:dm_av}}
	\end{table}
	
	\subsection{Comparisons to additional forecasting methods}
	
	Table \ref{tab:nz} presents additional empirical comparisons to a Gaussian model under the assumption of a zero mean ($\mu = 0$), and to quantile-based combinations of survey forecasts (`quantile mean' and `quantile median'). See the last two paragraphs of Section 7.2 in the main paper for details. 
	
	\begin{table}
		\centering
		\begin{tabular}{lccc}
			Model & Coverage & PI length & Interval Score \\ \toprule
			&&&\\
			\multicolumn{4}{c}{	\textit{German GDP (1991-2022, $n = 1307$)}}\\ \midrule		
			Gaussian & 79.11\% & 2.71 & 5.81 \\
			Gaussian ($\mu = 0$) & 78.50\% & 2.74 & 5.84 \\\bottomrule
			&&&\\
			\multicolumn{4}{c}{\textit{US GDP (1981-2022, $n = 320$)}}\\\midrule
			Gaussian & 76.56\% & 2.30 & 4.11 \\
			Gaussian ($\mu = 0$) & 78.12\% & 2.30 & 4.09 \\ \midrule
			SPF Histogram & 85.94\% & 2.97 & 4.48 \\
			SPF Histogram (quantile mean) & 79.38\% & 2.35 & 4.58 \\
			SPF Histogram (quantile median) & 77.50\% & 2.15 & 4.51 \\	
			\bottomrule
			&&&\\
			\multicolumn{4}{c}{\textit{US Inflation (1981-2022, $n = 320$)}} \\\midrule
			Gaussian & 78.44\% & 1.30 & 2.65 \\
			Gaussian ($\mu = 0$) & 77.81\% & 1.37 & 2.67 \\ \midrule
			SPF Histogram & 85.94\% & 2.25 & 3.35 \\
			SPF Histogram (quantile mean) & 81.88\% & 1.88 & 3.19 \\
			SPF Histogram (quantile median) & 80.00\% & 1.68 & 3.28 \\	
			\bottomrule
			&&&\\ \vspace{-.6cm}
		\end{tabular}
		\caption{The setup corresponds to Table 3 in the main paper, from which the `Gaussian' and `SPF Histogram' results are copied for easier reference. `Gaussian ($\mu = 0$)' denotes a restricted variant of the Gaussian method. The two additional SPF methods (US data only) represent different survey aggregation schemes as described in the text.\label{tab:nz}}
	\end{table}

\end{appendix}

\end{document}